\documentclass[manuscript]{aastex63}
\usepackage{color}

	\newcommand{\del}[1]{\partial_#1}
	\newcommand{\Dr}[1]{\frac{d}{d#1}}
	\newcommand{\DDr}[1]{\frac{d^2}{d{#1}^2}}
\received{}
\revised{}
\accepted{}
\submitjournal{ApJ}

\shorttitle{Gravitational waves from white dwarf merger remnants}
\shortauthors{Yoshida}

\begin{document}

\title{Decihertz gravitational waves from double white dwarf merger remnants}

\correspondingauthor{Shin'ichirou Yoshida}
\email{yoshida@ea.c.u-tokyo.ac.jp}

\author{Shin'ichirou Yoshida}
\affiliation{Department of Earth Science and Astronomy \\
Graduate School of Arts and Sciences, The University of Tokyo \\
Komaba 3-8-1, Meguro-ku, Tokyo 153-8902, Japan}



\begin{abstract}

Close binaries of double white dwarfs (DWDs) inspiral and merge
by emitting gravitational wave (GW). Orbital motion of some of these binaries are expected
to be observed at low frequency band by space-borne laser interferometric
detectors of GW. The merger remnant may suffer thermonuclear runaway
and explode as type Ia supernova if they are massive enough. As GW
sources the remnants have so far been scarcely studied. Here we propose a new mechanism
of GW emission from DWD merger remnants which may be observed
by planned GW detectors in decihertz frequency band. A remnant
is temporarily expected to have a high degree of differential rotation 
as a consequence of merger process. It is then unstable to oscillation
modes whose azimuthal pattern speed coincides with the stellar rotation.
We solve eigenvalue problem of differentially rotating remnants and
identify unstable eigenmodes which may be categorized to inertial modes.
The estimate of characteristic strain of GW shows that they 
may be detectable within the distance of the Virgo cluster
by planned gravitational wave observatories targeting the decihertz band.
\end{abstract}

\keywords{white dwarfs --- binary stars --- gravitational wave sources}


\section{Introduction} \label{sec:introduction}
%
Close binary systems composed of two white dwarfs, double white dwarf (DWD) binaries, emit
gravitational wave and may merge
within the age of the Universe. The gravitational wave from those inspiraling binaries
are expected to be seen (not necessarily welcomed, though, since it may be a hindrance to observe
other sources) 
by the proposed space-borne laser interferometric gravitational observatory
such as Laser Interferometer Space Antenna (LISA,~\cite{amaroseoane2012elisa}) or 
TianQin \citep{Luo_2016, PhysRevD.102.063021}.
The outcomes of the mergers may depend on the masses of the progenitor binaries. One of the most
interesting possibility is the merger of a binary whose total mass is close to or exceed the Chandrasekhar
mass may lead to type Ia supernovae (SNIa) of thermonuclear origin \citep{1984ApJ...277..355W, 1984ApJS...54..335I}.
The systems with the smaller total mass may survive the nuclear runaway after the merger
and may leave R Coronae Borealis stars \citep{1984ApJ...277..355W} or massive/strongly magnetized
white dwarfs \citep{2015SSRv..191..111F}.
The event rate for these mergers are crudely estimated as follows. 
Taking the merger rate $R_{\mbox{merge}}$ of double white dwarfs population 
in the Galactic disk to be $(6.3\pm 1.0)\times 10^{-13}\mbox{yr}^{-1}M_\odot^{-1}$ \citep{Maoz2018}
and extrapolated number density $n$ of Milky Way type galaxy to be $(1-1.5)\times 10^{-2}\mbox{Mpc}^{-3}$\citep{Kalogera2001},
we have the merger rate ${\cal R}$ of double white dwarf binary up to the distance $D$ as,
\begin{equation}
	{\cal R}(D) \sim \frac{4\pi}{3} D^3 R_{\mbox{merge}} \cdot n \cdot M_{\mbox{disk}} 
	\sim 14~\left(\frac{D}{20\mbox{Mpc}}\right)^3~\mbox{yr}^{-1}	
\end{equation}
Here Milky Way's disk mass $M_{\mbox{disk}}$ is taken from \cite{2015ApJ...806...96L}. Notice that this
rate takes into account neither the halo/bulge of disk galaxies nor the contribution from elliptical or dwarf galaxies.
\footnote{\cite{PhysRevD.102.063021} estimates the rate of
	inspiraling DWD with the total mass larger than $2M_\odot$
	in the Local Group of galaxies. Their event rate is
	smaller than obtained here, since the masses of
	the components are higher and the volume observed
	is smaller than are considered here.}

According to the hydrodynamical simulations of the merger, the less massive secondary star fills its Roche lobe
before the stars contact and the dynamical accretion of tidally disrupted secondary onto the massive
primary star occurs \citep{Benz1990,Segretain1997,Loren-Aguilar2005,Loren-Aguilar2009, 
Yoon2007,Dan2011,Dan2012,Pakmor2012,Raskin2012,Tanikawa2015,Sato2015,Sato2016}. Then
after the dynamical timescale the remnant settles into a rapidly and differentially rotating quasi-equilibrium state. It may evolve 
towards the shear-free state in a viscous timescale of typically $10^3-10^4$s \citep{Shen2012}. In this
paper we are interested in this quasi-equilibrium transient state of remnant, especially in the outcome of the large
degree of differential rotation.

Rotating stars are known to suffer a variety of instabilities which are driven by different physical mechanisms. 
An instability driven by such dissipative mechanism as viscosity, thermal transport or gravitational radiation is
referred to as secular instability, for which the timescale is determined by the characteristic time of the
particular dissipation mechanism\citep{Tassoul1978book}. Usually the timescale is longer than the
hydrodynamical one. Another category of instability, i.e., dynamical instability, has a purely hydrodynamic origin
and the growth time is the hydrodynamical timescale of the system under consideration. Classical studies
of self-gravitating homogenous spheroids/ellipsoids revealed that a spheroidal configuration is susceptible to a
dynamical instability
for $m=2$ perturbation ($m$ is the azimuthal wave number ) when it rotates above a critical speed. The rotation is conventionally
parametrized by $T/W$ value, where $T$ and $W$ are rotational kinetic energy and gravitational binding energy of a star respectively. 
The instability (bar-mode instabilty) sets in at $T/W\sim 0.27$ \citep{ChandraEFE, Tassoul_Ostriker1968, Ostriker_Tassoul1969, Ostriker_Bodenheimer1973, Tassoul1978book}.
The mechanism of it is explained as two particular oscillation modes merging into a pair of complex conjugate ones \citep{Schutz1980_pap3}.

Relatively recently by numerical simulations rotating stars with large degree of differential rotation have been found to suffer another dynamical instability \citep{Picket_etal1996, Centrella_etal2001, Shibata_etal2002, Shibata_etal2003, Saijo_etal2003, Ou_Tohline2006, Corvino_etal2010}. The instability is characterized by significantly lower threshold value of $T/W$
parameter than that of classical bar-mode, which may be as low as ${\cal O}[10^{-2}]$. The mechanism
of the instability has been explained as a corotation of wave pattern of unstable eigenmode with the stellar rotation
\citep{Watts_etal2005, Saijo_Yoshida2006, Passamonti_Andersson2015, 2016PhRvD..94h4032S, 2017MNRAS.466..600Y}, which explains
why the instability appears in rotating stars with large degree of differential rotation.

Though the saturation amplitude of the instability seen in hydrodynamical simulations is smaller than that of the classical bar-mode, it persists for hundreds of rotation periods
without being disrupted by non-linear hydrodynamic effects\citep{Shibata_etal2002, Saijo_Yoshida2006}. 
This characteristics of the instability is advantageous for detection of gravitational wave from it, once a compact star is susceptible to the instability.

In this paper linear eigenmode analysis is done for remnant models of binary white dwarf mergers in the early stage of their 
evolution. Unstable eigenmodes are
obtained for various mass, temperature and rotational parameter. The sequence of the fastest growing eigenmodes seems
to share common characteristics for all the models considered here. Moreover the gravitational wave strain from the unstable
oscillations are computed. We discuss the possibility of observing these gravitational wave transients by the proposed
detectors in ${\cal O}[10^{-2}] - {\cal O}[10^{-1}]$ Hz range.

\section{Formulation}

	\subsection{Equation of state}
	As for the equation of states of dense matter at finite temperature, we use a simple
	model of partially degenerate electrons with radiation.
	The number density of electrons $n_e$ with chemical potential $\mu$ 
	and temperature $T$ is written as \citep{1968pss..book.....C},
	\begin{equation}
		n_e= \frac{8\pi\sqrt{2}}{h^3} (m_e c)^3 \theta^{3/2} 
		\left(F_{1/2}(\eta, \theta)+\theta F_{3/2}(\eta,\theta)\right),
		\label{eq:electron number density}
	\end{equation}
	where $\eta \equiv \mu/m_ec^2$ and $\theta \equiv k_BT/m_ec^2$
	are a degenaracy parameter and a normalized temperature.
	Here $h, m_e, c, k_B$ are the Planck's constant, the mass of electron, 
	the speed of light in vacuum and Boltzmann constant.
	The functions $F_k(\eta, \theta)$ are Fermi-Dirac integrals defined as,
	\begin{equation}
		F_k(\eta, \theta) = \int_0^\infty \frac{x^k\left(1+\frac{\theta}{2}x\right)^{1/2}}{1+e^{x-\eta}} dx
		\label{eq:Fermi-Dirac integrals}
	\end{equation}
	where $k$ is the order of the integrals. \footnote{We compute the Fermi-Dirac integrals by using publicly available
	code sets from \url{http://cococubed.asu.edu/code\_pages/fermi\_dirac.shtml}.}
	Mass density of the gas $\rho$ is written as,
		\begin{equation}
			\rho = \left[\frac{A}{Z}\right]m_H n_e,
		\end{equation}
		where $m_H$ is the atomic mass unit. $\left[\frac{A}{Z}\right]$ is the mean ratio of atomic mass number $A$ to
		atomic number $Z$. Here we assume it to be two for simplicity. 
	Electron pressure $p_e$ is expressed as
	\begin{equation}
		p_e = \frac{16\pi\sqrt{2} m_e^4 c^5}{3 h^3}
		\theta^{5/2}\left(F_{3/2}(\eta, \theta)+\frac{\theta}{2} F_{5/2}(\eta,\theta)\right).
		\label{eq:electron pressure}
	\end{equation}
	Internal energy density $u_e$ is similarly written as
	\begin{equation}
		u_e = \frac{8\pi\sqrt{2} m_e^4 c^5}{3 h^3}
		\theta^{5/2}\left(F_{3/2}(\eta, \theta)+\theta F_{5/2}(\eta,\theta)\right).
		\label{eq:internal energy}
	\end{equation}
	Then the entropy of gas per mass $s_e$ is,
	\begin{equation}
		s_e = \frac{u_e+p_e}{\rho T} - \frac{\eta n_e k_B}{\rho}.
	\end{equation}
	
	We assume a simple isentropic model of a remnant, for which the specific entropy $s_e$ is kept constant in it.
	Then the adiabatic exponent $\Gamma$ which is necessary for the linear perturbation
	analysis is computed as follows. 
	Variating $s_e$ and set it zero, we have a linear adiabatic relation between $\Delta\eta$ and $\Delta\theta$
	where $\Delta$ mean small variations of $\eta$ and $\theta$,  This relation is used to evaluate adiabatic
	exponent of the gas $\Gamma_e$,
		\begin{equation}
			\Gamma_e \equiv \frac{\Delta \ln p_e(\Delta\eta, \Delta\theta)}{\Delta\ln\rho(\Delta\eta, \Delta\theta)}.
		\end{equation}

	Photon's radiation pressure $p_\gamma$ is added to $p_e$ to produce the total pressure,
	\begin{equation}
		p_\gamma = \frac{a}{3}T^4,
	\end{equation}
	where $a$ is the radiation density constant.
	
	The total adiabatic exponent $\Gamma$ is then given by \citep{1978trs..book.....T}
	\begin{equation}
		\Gamma = \beta_e + \frac{(\Gamma_e-1)(4-3\beta_e)}{\beta_e+12(\Gamma_e-1)(1-\beta_e)},
		\label{eq:total adiabatic exponent}
	\end{equation}
	where $\beta_e = p_e/(p_e+p_\gamma)$. As is easily seen, $\Gamma \to 4/3$ as $\beta_e \to 0$.

	\subsection{Equilibrium models}
	To compute the remnant models, we assume them to be in dynamical equilibrium with
	axisymmetry. Isentropic nature of the configuration makes it possible to apply Hachisu's
	self-consistent field method \citep{1986ApJS...61..479H}. With a prescribed angular
	frequency profile $\Omega = \Omega(r\sin\theta)$, where $r$ and $\theta$ are the polar
	radial and angular coordinates, we may compute equilibrium models of rotating stars
	up to mass-shedding limit.
	\footnote{By Poincar\'{e}-Wavre theorem, the angular frequency depends only on the distance
	from the rotational axis \citep{1978trs..book.....T}}.
	Rotational deformation is conventionally measured by the axis ratio parameter $r_p/r_e$
	where $r_p$ and $r_e$ are the polar and the equatorial radius of the model.
	
	As for the rotational profile we adopt the differential rotation law
	in \cite{2019MNRAS.486.2982Y}, especially called 'Yoon07' there, which analytically
	fit the rotational profile of a merger simulation in \cite{2007MNRAS.380..933Y}. The
	profile has a slowly and uniformly rotating core and a nearly Keplerian envelope.
	The explicit expression is
	\begin{equation}
		\frac{\Omega(R)}{\Omega_c} = \beta\left(
		\left(1+\left(\frac{R}{R_c}\right)^{3/2}\right)^{-1}(1+e^{-\alpha x})^{-1}
		- (1+e^\alpha)^{-1}
		\right) + 1 ; \quad x \equiv 2r^B-1,~ B = \frac{\ln\frac{1}{2}}{\ln R_c}
		\label{eq: Omega profile}
	\end{equation}
	The profile depends on four parameters, $\Omega_c, R_c, \alpha, \beta$.
	They are respectively, the angular frequency at the origin, 
	the radius of the uniformly rotating core normalized by the equatorial radius
	of the star, the parameter
	controlling the steepness of the $\Omega$ increment above the core,
	and the parameter controlling the maximum ratio of $\Omega$ to
	$\Omega_c$.
		\begin{figure}[htpb]
		\centering
		\includegraphics[scale=0.5]{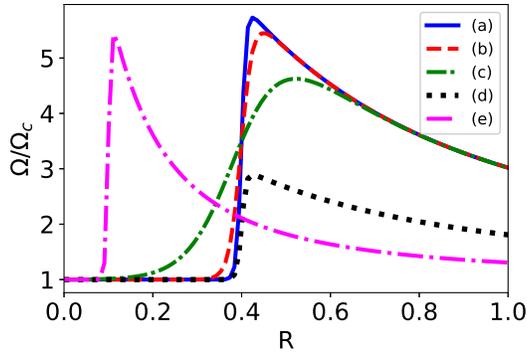}
		\caption{Profiles of angular velocity. The radial coordinate $R$
		is normalized by the equatorial radius. The angular frequency $\Omega$ is normalized by
		the value at the origin $\Omega_c$. (a) $(\alpha,\beta,R_c)=(100,10,0.4)$.
		(b) $(\alpha,\beta,R_c)=(40,10,0.4)$. (c) $(\alpha,\beta,R_c)=(10,1,0.4)$. (d) $(\alpha,\beta,R_c)=(100,4,0.4)$. 
		(e) $(\alpha,\beta,R_c)=(100,10,0.1)$.}
		\label{fig: eq-Omega}
		\end{figure}
In Fig.\ref{fig: eq-Omega} we show typical angular frequency profiles for different
sets of the parameters. Notice that $\Omega$ above the core asymptotes to the
Keplerian distribution $\Omega \propto R^{-3/2}$.
	
	We fix the temperature and the density at the center to compute the specific entropy
	of the models. The thermodynamic quantities in the stellar interior are computed by
	assuming constancy of the entropy.

	\subsection{Perturbations\label{sec: formulation-perturbation}}
	In this study we perform linear perturbation analysis of the remnant models.
	The formulation and the numerical code are based 
	on \cite{Saijo_Yoshida2006, 2017MNRAS.466..600Y}
	where the equilibrium star is approximated as a flattened configuration. Then
	we solve a set of ordinary differential equations of radial coordinate. 
This approximation is to make the problem largely simplified. Otherwise
we would need to solve the two-dimensional eigenvalue problem whose computational
domain is the meridional section of a star. This is rather hard a problem especially
when we have corotation radii at which the stellar rotation coincides with the
azimuthal pattern speed of an eigenmode (see Eq.(\ref{eq:2potentials_dPhi}) and the comment
below the equation).
We may justify the formulation if the eigenmodes here share the same
characteristics as the eigenmodels susceptible to the low T/W instability.
In \cite{Saijo_Yoshida2006} we found the characteristics of low-T/W instability
using three-dimensional hydrodynamical simulations.
The panel II of Fig.5 in \cite{Saijo_Yoshida2006}
shows the characteristic velocity pattern of low T/W instability
seen in the simulation. It is noticed that the velocity perturbation
lacks the component parallel to the rotational axis anywhere
except at the surface. This planar velocity perturbation
means that the dimensionless density perturbation 
also has a planar profile perpendicular to the axis.
Then the perturbation is quasi-planar and the equations
may be well-approximated by the radial ordinary differential equations
when the azimuthal dependence is decomposed into harmonics.
Checking the validity of the approximation is beyond our scope here.
Thus the results shown below are presented to show a 'proof of principle'
and should not to be taken as the detailed results for realistic remnants.

Our "cylindrical model" deals with a system of equation only in the equatorial plane and neglect the $z$-dependence 
of the perturbed variables.  
We utilize the two potential formalism of \cite{1991ApJ...379..285I}. The linear adiabatic perturbation equations for 
a stationary and axisymmetric fluid configuration are cast into a coupled system of two scalar potentials; $\delta \Phi$, 
the perturbed gravitational potential and $\delta U$, the sum of the perturbed enthalpy and the gravitational potential.
Both of the potentials are defined in the Eulerian description of fluid.
In cylindrical coordinate $(R,\varphi,z)$ we have two equations of the potentials as
	\begin{eqnarray}
	\label{eq:2potentials_dU}
		&&0 = \DDr{R}\delta U\nonumber\\ 
		&+& \left[\del{R}\ln\left(\frac{s}{L}\right)-\frac{2m\Omega}{sR}+\frac{1}{R}
			+\frac{\del{R}\rho}{\rho}+\frac{m\kappa^2}{2\Omega sR}\right]\Dr{R}\delta U \nonumber \\
	&& + \left[-\frac{L}{s}\del{R}\left(\frac{2m\Omega}{RL}\right)+\frac{L}{c_s^2}
			-\frac{2m\Omega}{sR}\left(\frac{1}{R}+\frac{\del{R}\rho}{\rho}\right)-\frac{m^2}{R^2}\right]\delta U\nonumber \\
			&&- \frac{L}{c_s^2}\delta\Phi,
	\end{eqnarray}
and
	\begin{equation}
			\label{eq:2potentials_dPhi}	
		\left(\DDr{R}+\frac{1}{R}\Dr{R}-\frac{m^2}{R^2}\right)\delta\Phi = 4\pi\rho\left(\frac{d\rho}{dp}\right) (\delta U-\delta\Phi),
	\end{equation}
	where we assume harmonic dependence of Eulerian perturbations on $t$ and $\varphi$ as $\sim e^{-i\sigma t+im\varphi}$.
Here $\rho$ is the equilibrium density, $\Omega$ is the angular velocity of equilibrium flow, $\kappa^2= 2\Omega\left(2\Omega + R\del{R}\Omega\right)$ is the epicyclic frequency squared, $s=\sigma-m\Omega$ is the frequency seen from a co-moving observer to the equilibrium flow, $L=s^2-\kappa^2$ and the sound speed of equilibrium fluid is $c_s$. 
We call the zero of $s$ as a corotation singularity, which is a simple pole
of equation (\ref{eq:2potentials_dU}) in the complex $R$-plane.
In the former studies \cite{Saijo_Yoshida2006,2017MNRAS.466..600Y} 
we assume the imaginary part of $\sigma$ is much smaller than the real part. Then 
the zero of $s$ is regarded to locate on the real $R$-axis. The path bypassing the zero
in the complex $R$-plane leads to the small imaginary part of the eigenfrequency.
In this work we do not assume the imaginary part to be much smaller than the real
part, thus the integration path of the perturbation equation on the real $R$-axis
do not have a singularity. 

The search for an eigenvalue is done as follows. For a given complex number $\sigma$, we numerically
integrate the equations from the center and the surface of the star. The boundary
conditions imposed are the same as in \cite{Saijo_Yoshida2006,2017MNRAS.466..600Y},
i.e., the regularity at the center and the free surface condition as well as the regularity
of the gravitational potential at the infinity. At a matching point inside the star we compute
the Wronskian of the solutions. The zeroes of the Wronskian are eigenvalues
and we search for them by using Muller's method \citep{1992nrfa.book.....P}
in the complex $\sigma$-plane.

\subsection{Gravitational wave}
\subsubsection{Quadrupole formula}
To compute the gravitational wave amplitude we adopt
the quadrupole formula of the metric function for fluid systems
\citep{1989PThPh..82..535O,1990MNRAS.242..289B,1997A&A...320..209Z}.
Transvers-Traceless (TT) component $h_{ij}^{TT}$ of the spatial part
of the metric perturbation (by which full metric is written as $\eta_{\mu\nu} + h_{\mu\nu}$
with $\eta_{\mu\nu}=\mbox{diag}(-1,1,1,1)$) is\footnote{Here Latin indices are for spatial coordinates, 
while Greek ones are for spacetime coordinates.}
	\begin{equation}
		h_{ij}^{TT} = \frac{2G}{c^4D}{\bf P}_{ijk\ell} \int d^3\vec{x} \rho\left[2 v^kv^l - x^k\partial^\ell\Phi
		-x^\ell \partial^k\Phi \right],
		\label{eq: hTT}
	\end{equation}
where $\rho$, $v^k$, $\Phi$, $D$ are mass density, velocity of fluid, gravitational potential, and
the distance to the source. 
Here the cartesian
coordinate is used whose origin is at the center of mass of the gravitational wave source.
Notice that $\partial^k=\partial_k$ for the cartesian coordinate.
${\bf P}$ is the projection tensor perpendicular to the line of sight to the source,
	\begin{equation}
		{\bf P}_{ijk\ell} = \gamma _{ik}\gamma_{j\ell} - \frac{1}{2}\gamma_{ij}\gamma_{k\ell},
		~ ; \gamma_{ij}\equiv \delta_{ij} - N_iN_j
	\end{equation}
where $\delta_{ij}$ is the unit tensor and $N_i$ is the directional unit vector pointing towards
the observer from the origin. The viewing angle is expressed by $N_i$ which we choose
to coincide with the unit vector parallel to the rotational axis of the star (i.e., z-direction)
for simplicity.

We here focus on the quadrupole mode of stellar oscillation with the azimuthal wave number
$m=2$, thus the eigenfunctions linearly enter to Eq.(\ref{eq: hTT}).

It should be remarked that the spatial integration in Eq.(\ref{eq: hTT}) is performed by assuming
the planar behavior perpendicular to the rotational axis, as is remarked in Sec.\ref{sec: formulation-perturbation}.

\subsubsection{Characteristic strain\label{sec:strain}}
For such a gravitational wave source with a secularly changing frequency as an inspiraling compact
binary, the characteristic strain $h_c$ is the conventional signal parameterization of gravitational wave
(see e.g., \cite{2015CQGra..32a5014M}). It is related to the number of cycles $N_c$ for which the oscillations
of gravitational wave strain with a given frequency lasts, 
	\begin{equation}
		h_c = \sqrt{N_c} h_0, 
		\label{eq: hc}
	\end{equation}
where $h_0$ is the amplitude of nearly sinusoidal gravitational wave strain.

Eigenfrequencies of a merger remnant changes as it secularly evolves in a secular
time scale longer than the dynamical one. 
The viscous timescale is the second shortest among the
relevant timescales, i.e., dynamical, viscous, and thermal.
It is the one during which the rotational profile
of the remnant changes and the change affects the eigenfrequency.
The viscous timescale is estimated for two cases in \cite{Shen2012}. 
One of them assumes the $\alpha-$viscosity \citep{1973A&A....24..337S} resulting
from magneto-rotational turbulence. The order of timescale is $10^{4}$ sec. It should, however, be noticed that the criterion
of the magneto-rotational instability \citep{Velikhov1959, 1960PNAS...46..253C, Balbus-Hawley1991a} 
is not satisfied at the core-envelope boundary, where $d\Omega/dr>0$. 
The other one comes 
from Tayler-Spruit dynamo \citep{Tayler1973, 1999A&A...349..189S, 2002A&A...381..923S, 2008ApJ...679..616P},
which results in the timescale of $10^{4}$ seconds for our typical rotational profile ((a) in Fig.\ref{fig: eq-Omega}).
For the estimate above, we use Eq.(1) and Eq.(5) in \cite{Shen2012}.
Overall the viscous timescale in which differential rotation
of remnants is smoothed out may be $10^4$ s.
The duration of saturated oscillations is expected to be shorter than this
timescale, since it takes e-folding time for a unstable mode to
saturate. This e-folding time is typically less than $10^2$ s
for the fastest growing eigenmodes. I choose the typical
duration of saturated oscillation to be $10^3$ s.
Thus the number of cycles of the initially unstable oscillation with frequency $f$ is at most
$N_c = f\tau_0$. Here we fix $\tau_0$ to be $10^3$s for simplicity and compute
the characteristic strain by Eq.(\ref{eq: hc}).

\subsubsection{Fixing amplitude of the eigenmodes}
To estimate gravitational wave amplitude of eigenmode oscillations of stars, we need to specify
the amplitude of the eigenmode functions. This is not in principle done within the linear theory, for which
the amplitude is 'small' but arbitrary. We thus parametrize the amplitude of the modes.
\cite{Shibata_etal2002} studied $N=1$ polytrope with differential rotation by using nonlinear
hydrodynamical simulations and found that the low T/W dynamical instability has an amplitude
of ${\cal O}[10^{-2}\sim 10^{-1}]$ measured by a distortion parameter 
$\eta = |I_{xx}-I_{yy}|/(I_{xx}+I_{yy})$ where $I_{ij}$ are instantaneous moments of inertia.
While \cite{2009ApJ...701..225L} and \cite{2013MNRAS.430.1988M} argued that the Rossby-wave instability
in astrophysical disks may be saturated at the amplitude of velocity perturbation
for which the circulation time of the local vortex is comparable to the growth time of the
instability.

We here introduce the ellipticity $\epsilon_I$ of the deformed star due to the eigenmode
as
\begin{equation}
	\epsilon_I = \left|\frac{\delta I_{zz}}{I_{zz}}\right|
	= \left|\frac{\int dV \delta\rho R^4}{\int dV \rho R^4}\right|.
	\label{eq: ellipticity_I}
\end{equation}
The amplitude of the eigenmode scales linearly as $\epsilon_I$. 
As the parameter may be a proxy of the deformation parameter $\eta$ in \cite{Shibata_etal2002}, 
we may set $\epsilon_I\sim{\cal O}[10^{-2} - 10^{-1}]$. 
On the other hand when we apply the estimate of saturation amplitude of the Rossby-wave instability,
we obtain the amplitude of ${\cal O}[10^{-5}]$.  Therefore we here fix the range of the amplitude 
parameter to be $10^{-5}\le \epsilon_I \le 10^{-1}$.

\section{Results}
We here focus our attention to $m=2$ modes which are expected to be the most important
oscillations that couple to gravitational wave.
\footnote{$m=1$ modes may be excited in a merger remnant \citep{Kashyap2017},
but the coupling to gravitational wave is at the second order in small perturbation
in the quadrupole formula. Higher order than $m=2$ generally contributes to gravitational
radiation with smaller amplitude.
}
\subsection{Unstable eigenmodes}
\begin{figure*}
	\gridline{\fig{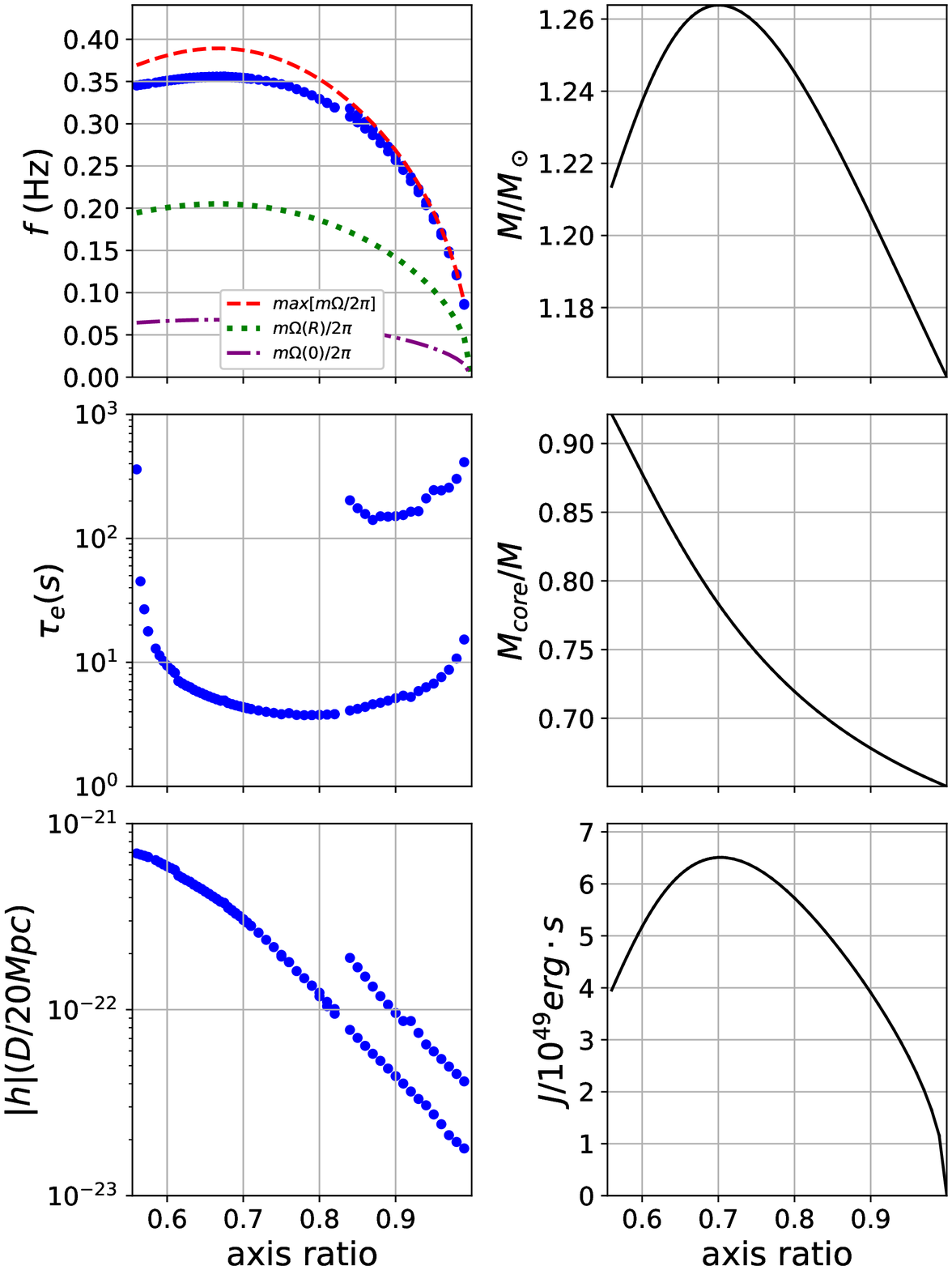}{0.4\textwidth}{(a): $R_c=0.4$}
			\fig{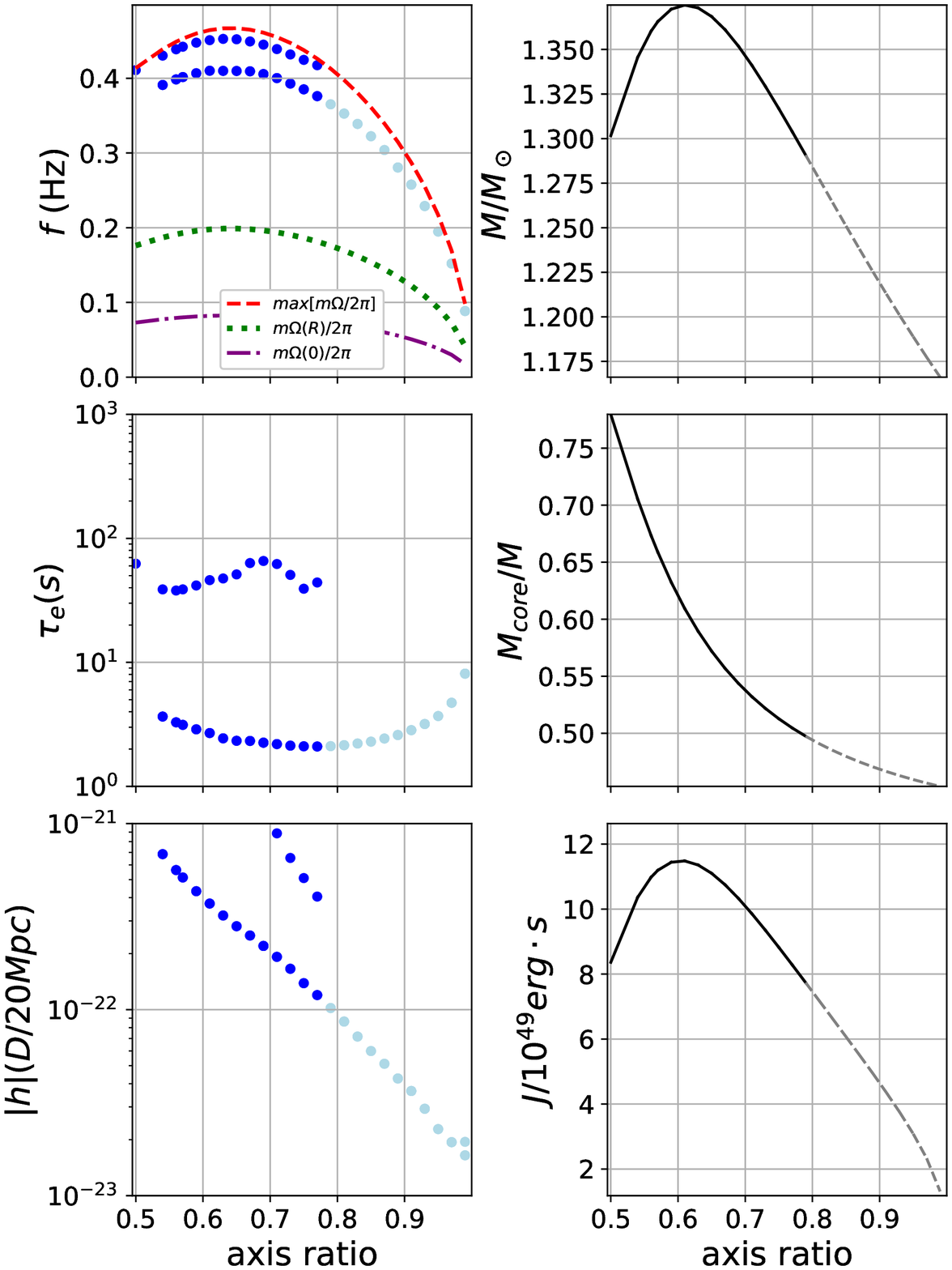}{0.4\textwidth}{(b): $R_c=0.3$}
			}
\caption{Sequences of eigenmodes and corresponding equilibrium models.  Common parameters
in both panels are $(T_c~({\rm K}), \rho_c({\rm g}{\rm cm}^{-3}), \alpha, \beta)=(10^7, 10^8, 100, 10)$.
The panels (a) are for $R_c=0.4$ while those of (b) are for $R_c=0.3$.
Top left: Real part of eigenfrequency $\Re[\sigma/2\pi]$ in Hz. Blue dots correspond to the eigenfrequency
while the solid curve is the maximum of the corotation frequency. The dotted and the dot-dashed curve are
the corotation frequency at the equatorial surface and at the origin, respectively.
Middle left: Growth time $\tau_e=\Im[\sigma]^{-1}$
of the eigenmodes. Bottom left: dimensionless strain $h$ of gravitational wave times the source distance $D$
in unit of $20$Mpc. Amplitude parameter $\epsilon_I=10^{-2}$. Top right: Total mass of the equilibrium remnant.
Middle right: Mass fraction of the core. Bottom right: Total angular momentum in units of $10^{49}$erg s.
The leftmost point corresponds to the mass-shedding limit.
In the panels (b) Thin dots and the dashed lines correspond
to the models whose mass fraction of the core is less than 0.5.
\label{fig: T7rho8Rc0403}}
\end{figure*}

In Fig.\ref{fig: T7rho8Rc0403} we plot eigenfrequency and equilibrium variables
as functions of axis ratio, which measures the rotational deformation of equilibria. Panels of (a)
are for $R_c=0.4$, while those of (b) are for $R_c=0.3$. Other parameters are the same
for both models, i.e., $(T_c~({\rm K}), \rho_c({\rm g}{\rm cm}^{-3}), \alpha, \beta)=(10^7, 10^8, 100, 10)$.
Notice that the smaller axis ratio corresponds to the larger deformation of the star from the sphere (when
the axis ratio is unity). First of all we need to be careful to interpret the equilibria as the
sequence parametrized by 'speed of rotation'. As the axis
ratio decreases from unity the equilibrium suffer the large degree of centrifugal distortion. Along
with it the angular momentum increases when the axis ratio is close to unity. 
However it is not a monotonic function of the axis
ratio. This results from the highly non-monotonic distribution of the angular velocity 
(see Fig.\ref{fig: eq-Omega}). It also follows that the total mass is not monotonic as is seen
in the top right panel. 

In both panels (a) and (b),
the sequence of the real part of eigenfrequency ($\Re[\sigma/2\pi]$) is plotted in the top left panel
(blue dots). The red solid curve is the maximum of the corotation frequency inside
the star, $\max\left[m\Omega(R)/2\pi\right]$. For these equilibrium sequences two discrete sequence
of unstable eigenmodes are specified. Both of the mode are within the corotation resonance
of frequency, $m\Omega(0)\le\Re[\sigma]\le{\rm max}[m\Omega]$. 
For $R_c=0.4$ case (panel (a)) frequencies of these modes asymptote to zero. One of the
mode is unstable all the way through the mass-shedding limit. This mode is the fastest growing
mode. The other mode seem to exit from the corotation resonance and be stabilized around
the axis ratio of 0.83.
The sequence of these modes, whose limiting frequency for non-rotating
star (axis ratio being unity) is zero, suggest that they are inertial type of oscillations
whose main restoring force is Coriolis force, rather than acoustic type. 
\cite{2020arXiv200310198P} recently reported that
remnant models of binary neutron star merger exhibit low T/W instability (see also \cite{2020PhRvD.102d4040X}).
They find there are two types of unstable modes, which corresponds to acoustic type (f-mode)
and the rotationally supported inertial mode (i-mode). Although their functional form of
the profile of the rotational angular frequency is different from ours, they share a qualitative
similarity, that is, nearly flat profile around the origin and the off-center maximum of $\Omega$.
It should be noted that the frequency of the unstable modes is close to the maxima of
corotaion frequency. We find no unstable modes whose corotation point is either close to
the origin or to the stellar surface.
Remarkably in the current study the fastest growing modes of equilibrium stars for the other parameter sets
share the same inertial type of nature.

For the case with $R_c=0.3$ (b) we see the equilibrium model may have the core fraction
less than 0.5. Our assumption on the remnant is that the secondary of the progenitor binary
is disrupted and accreted onto the primary, which forms the envelope
of the remnant while the core consists of the matter in the primary. As a result the core 
mass should be larger than the envelope mass. We therefore discard the portion of the sequence
with the mass fraction being less than 0.5. For the sequence in (b) the models with
the axis ratio less than 0.78 is the allowed models. We see the fastest growing mode
has the characteristics of the inertial mode (see the top left). The other less unstable mode
seems to have a different character. As the axis ratio decreases from unity (the degree of
rotational deformation increases), the mode which is initially outside the corotation band
(below the red dashed curve of the top left panel) seems to enter the corotation band
and becomes unstable. The mode seems not to asymptote to the zero frequency mode
and may have an acoustic nature.


\begin{figure*}
	\gridline{\fig{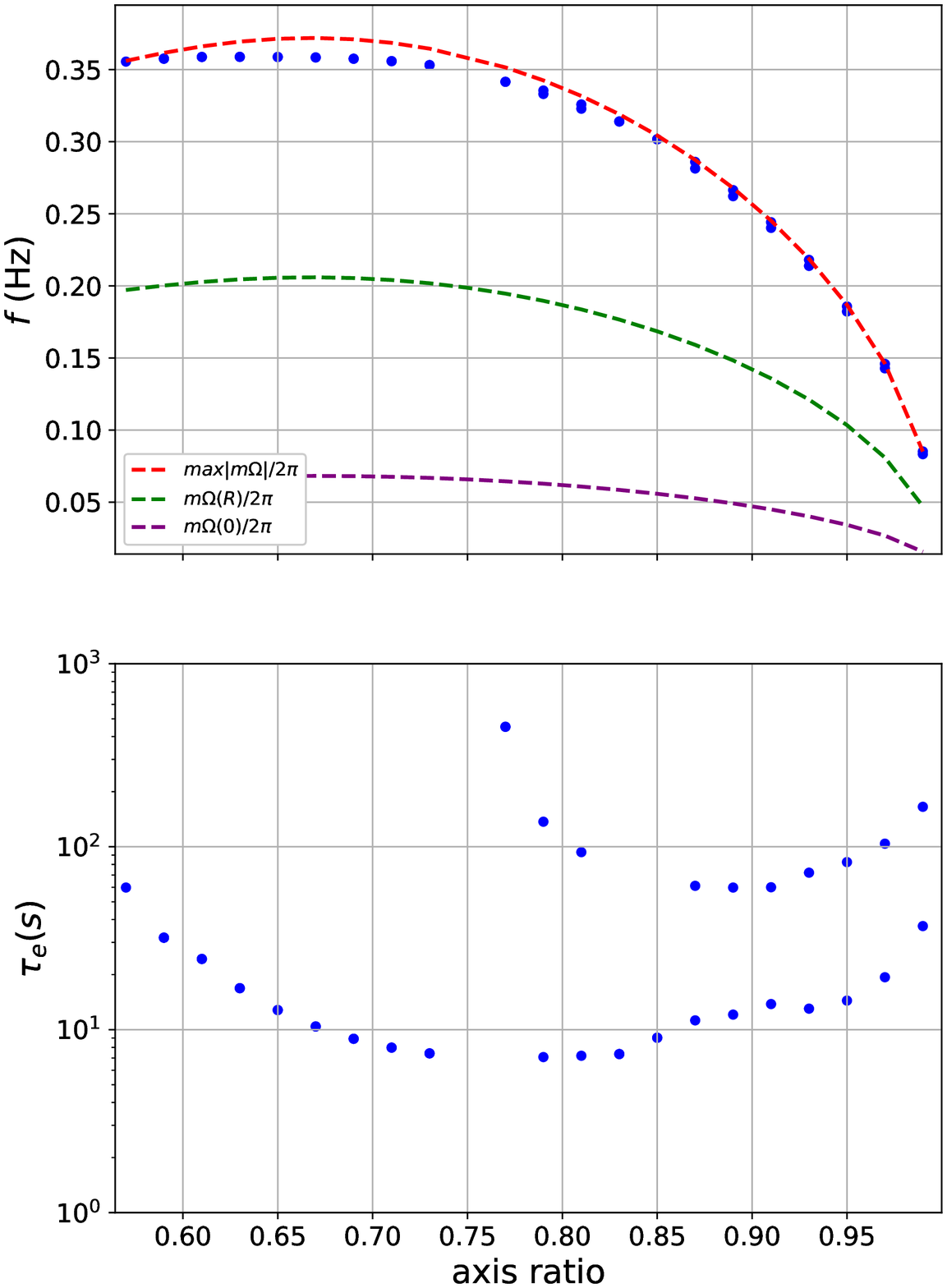}{0.3\textwidth}{(a)}
			\fig{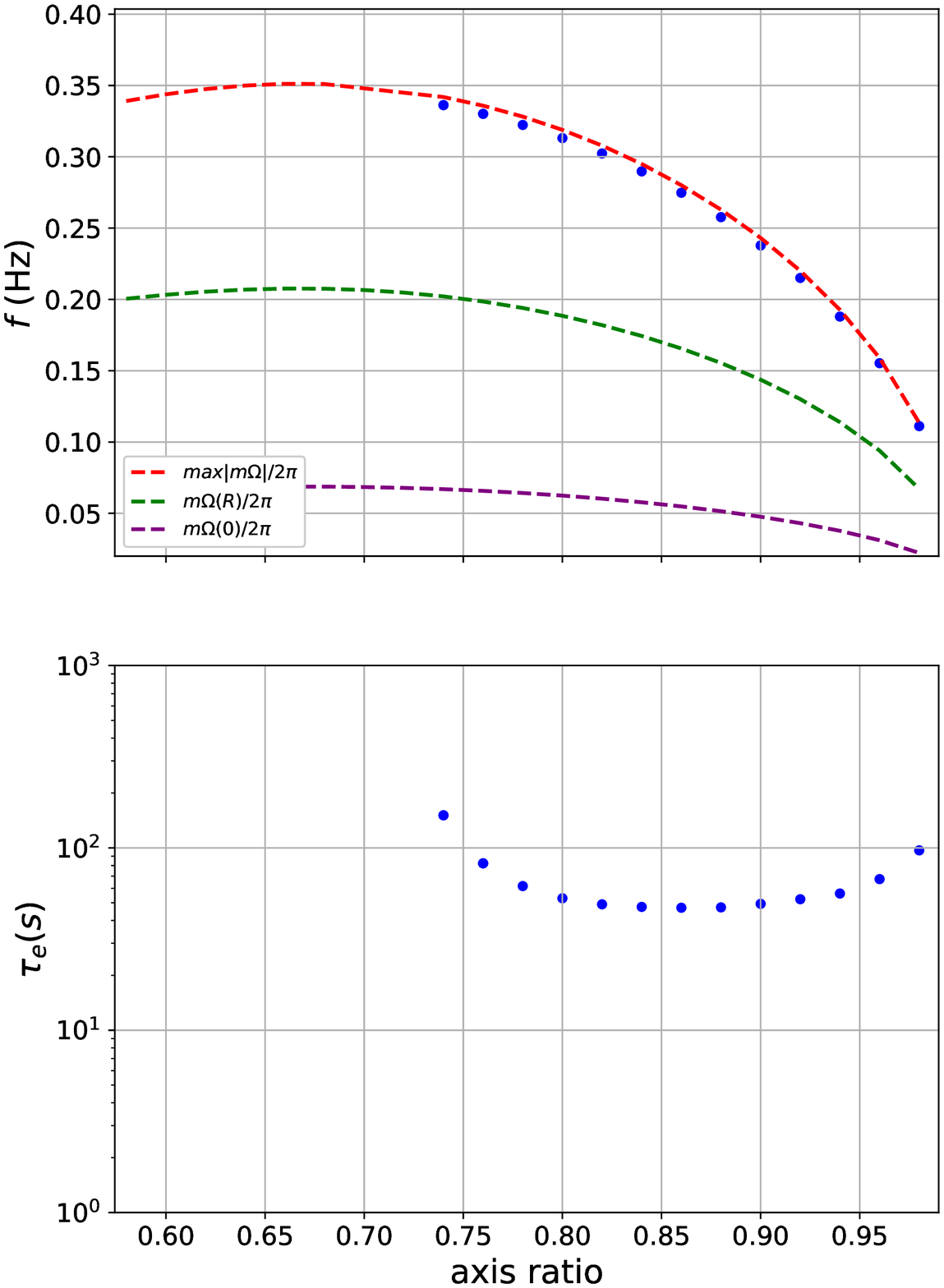}{0.3\textwidth}{(b)}
			\fig{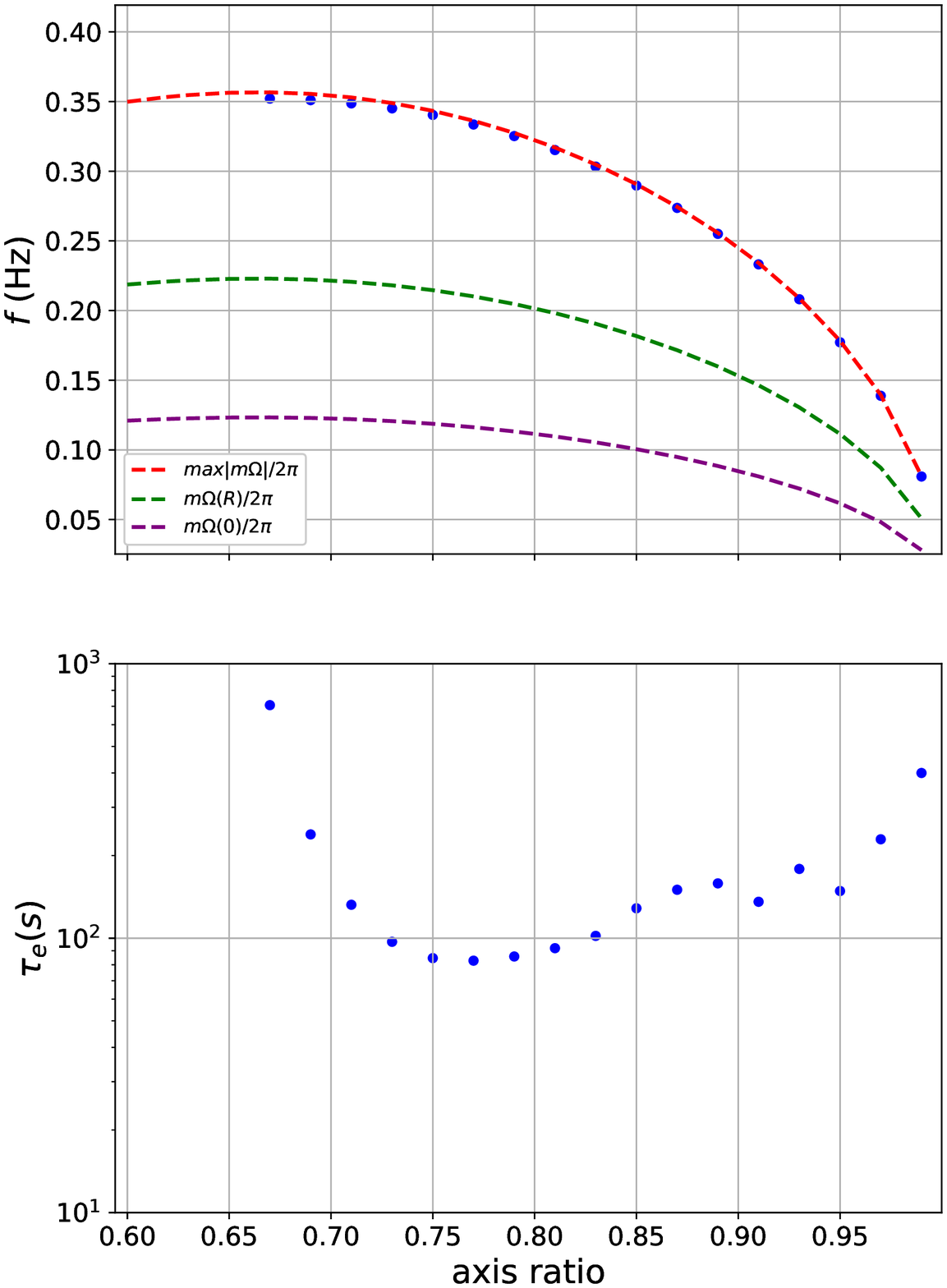}{0.3\textwidth}{(c)}
			}
	\caption{(a) Eigenfrequency $\Re[\sigma/2\pi]$ (top panel) and growth time $\tau_e=\Im[\sigma]^{-1}$
for  $(T_c, \rho_c, \alpha, \beta, R_c)=(10^7, 10^8, 40, 10, 0.4)$.
(b) $(T_c, \rho_c, \alpha, \beta, R_c)=(10^7, 10^8, 20, 10, 0.4)$ .
(c) $(T_c, \rho_c, \alpha, \beta, R_c)=(10^7, 10^8, 100, 4, 0.4)$.
\label{fig: modeT7rho8Rc0.40_bet10set}}
\end{figure*}
(c) is the model with smaller $\beta$ than in Fig.\ref{fig: T7rho8Rc0403}. $\beta$
is a measure of the maximum $\Omega$ as is seen in Fig.\ref{fig: eq-Omega}. For the value of $\beta\lesssim 2$,
we find no unstable mode. 

\begin{figure*}[htpb]
	\gridline{\fig{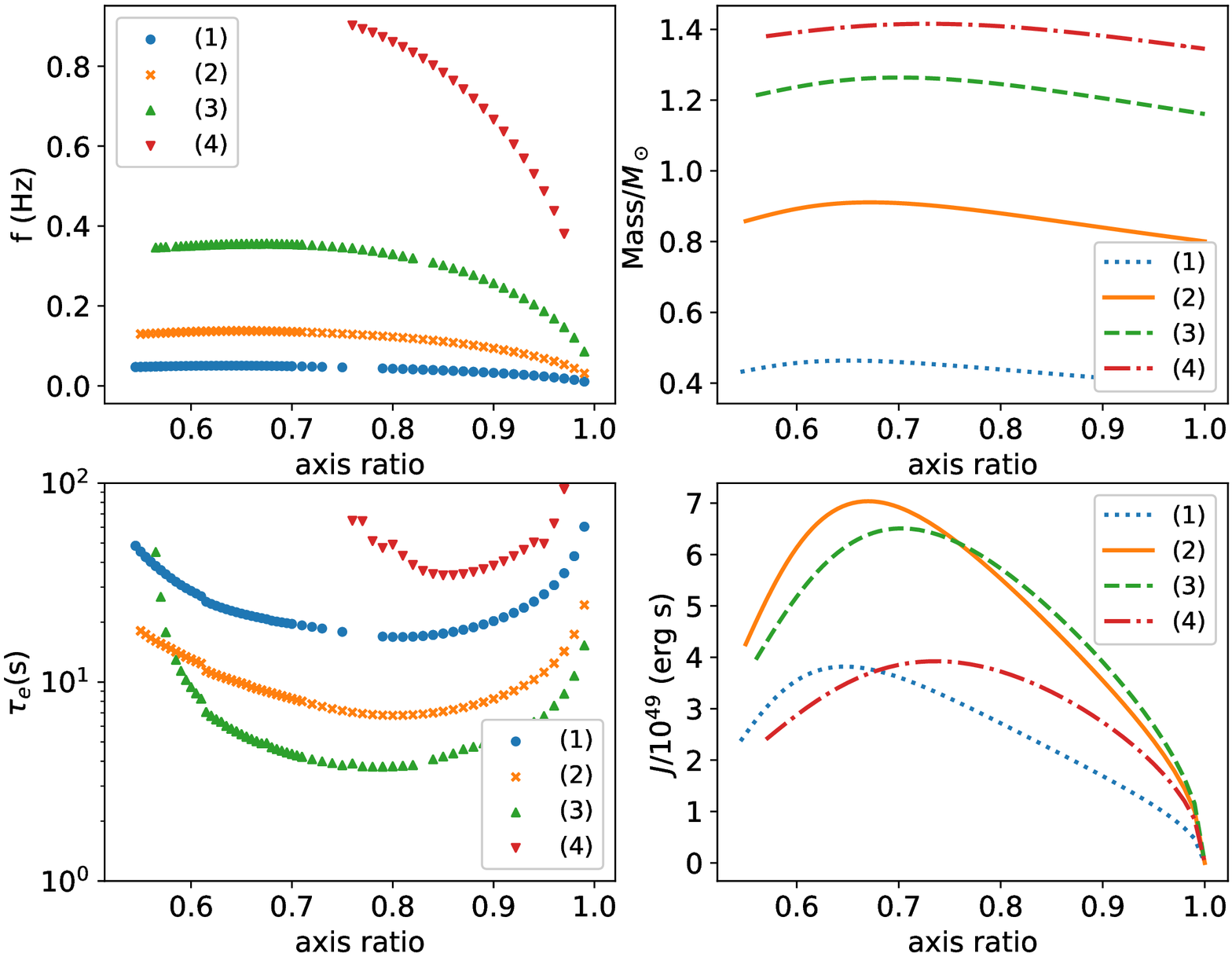}{0.47\textwidth}{(a)}
			\fig{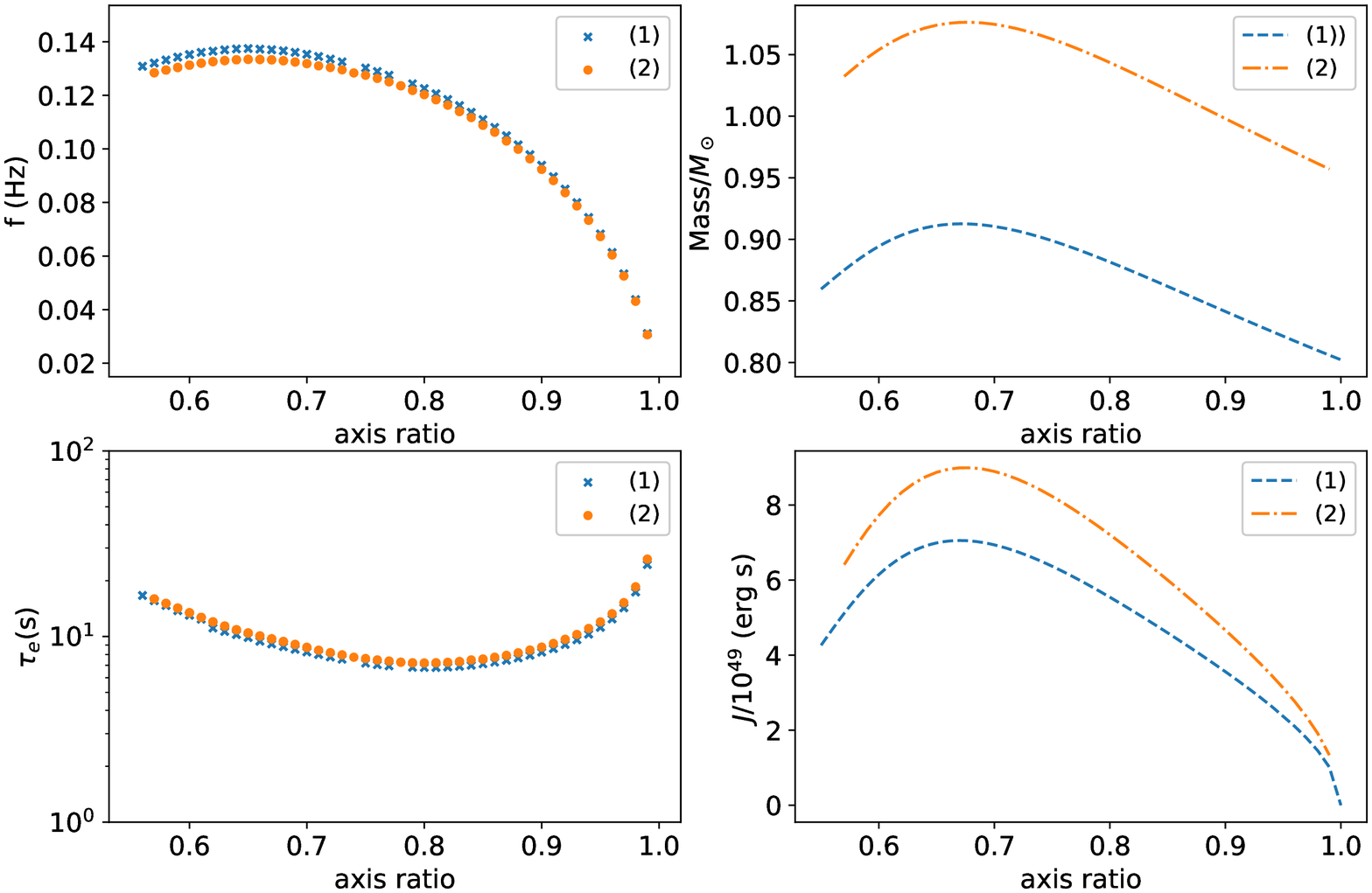}{0.53\textwidth}{(b)}
			}
	\caption{Eigenfrequency $\Re[\sigma/2\pi]$ (top left), growth time $\tau_e=\Im[\sigma]^{-1}$
	(bottom left), mass in $M_\odot$ (top right), and angular momentum in $10^{49}$ erg s.
	In the panels (a), the temperature $T_c$ is fixed to be $10^7$K and the central density $\rho_c$
	is varied. (1): $\rho_c=10^6{\rm g}{\rm cm}^{-3}$, (2): $\rho_c=10^7{\rm g}{\rm cm}^{-3}$, 
	(3): $\rho_c=10^8{\rm g}{\rm cm}^{-3}$, (4): $\rho_c=10^9{\rm g}{\rm cm}^{-3}$.
	In the panels (b), $\rho_c$ is fixed to be $10^7{\rm g}{\rm cm}^{-3}$, while $T_c$ is varied.
	(1): $T_c=10^8$K, (2): $T_c=10^9$K.
\label{fig: T-rho-comparison}}
\end{figure*}

Fig.\ref{fig: modeT7rho8Rc0.40_bet10set}
is compared with the top and middle left panels of Fig.\ref{fig: T7rho8Rc0403}.
(a) and (b) correspond to the smaller $\alpha$ models, for which the gradient of $\Omega$
at the core-envelope interface are smaller (see Fig.\ref{fig: eq-Omega}). When the gradient
of $\Omega$ becomes too small (when $\alpha\lesssim 10$) we find no unstable eigenmode.
Notice that this variation of $\alpha$ modifies very little such bulk quantities as the total mass, 
core mass fraction, and the angular momentum.



In Fig.\ref{fig: T-rho-comparison} we compare the equilibria ant the eigenmodes
for different $\rho_c$ (panels (a)) and for different $T_c$ (panel (b)).
$R_c, \alpha, \beta$ are fixed.
The unstable modes are of the inertial type. The eigenfrequency of the modes
is higher for the higher $\rho_c$ (thus massive) models. The growth time
is not a monotonic function of the density or the mass. rather It may be
correlated with the angular momentum of the star (the bottom right
of (a)). The model with higher angular momentum seems to have the
smaller growth time.

In the panels (b) we plot the eigenfrequency, the growth rate as well
as the mass and the angular momentum for different value of $T_c$.
The model (2) with $T_c=10^9$K correspond to the higher temperature and the mass
is larger due to the support of the thermal pressure as well as
the degenerate electron pressure. The model (1) has $T_c=10^8$K
for which thermal pressure makes small difference from the lower
temperature models. Only weak dependence on $T_c$
is seen both in the frequency and the growth time, at least as high
as $T_c\sim 10^9$K.

\subsection{Characteristic strain of gravitational wave from unstable modes}
One of the most interesting questions on the oscillations of the merger remnants is whether 
they are observable by some of the planned detectors of gravitational waves. Since the typical frequency
of the remnants are in the deci-Hz range, they are neither good targets for the now-operating
interferometers such as Laser Interferometer Gravitational-wave Observatory (LIGO, \url{https://www.ligo.org/}), 
Virgo (\url{https://www.virgo-gw.eu/}), and KAGRA (\url{https://gwcenter.icrr.u-tokyo.ac.jp/en/}), nor planned LISA, 
the space-borne gravitational wave detector for low-frequency range. The deci-Hz range gravitational wave detections
are currently planned with different technologies, because of such astrophysical/cosmological
interests as discovery of intermediate-mass black holes, cosmological background gravitational waves, 
precise measurements of neutron stars/black holes binary characteristics \citep{sedda2019missing}. 
One type of detectors is the space-borne antennae system similar
to LISA but with shorter baselines, i.e., Big Bang Observer (BBO, \cite{PhysRevD.72.083005}), DECi-hertz Interferometer 
Gravitationa-wave Observatory (DECIGO, \cite{Kawamura_2006}), and TianQin \citep{Luo_2016}.
Other types of detectors proposed are TOrsion Bar Antennae (TOBA, \cite{PhysRevLett.105.161101})
and atomic interferometers \citep{2017arXiv171102225G, 2020JCAP...05..011B, 2019arXiv190800802A}.

For simplicity we compute the characteristic strain of gravitational wave from dynamically unstable remnants 
by assuming the following conditions. First we assume only the fastest growing mode contributes
the strain. Secondly the saturated mode keeps its frequency and amplitude for viscous timescale $\tau_0$ 
(see Sec.\ref{sec:strain}) and diminishes.
Inclusion of the multiple unstable modes and the evolution of eigenmodes as the remnant evolves in viscous
timescale is beyond our scope here.

\begin{figure}[htpb]
\centering
\includegraphics[scale=0.6]{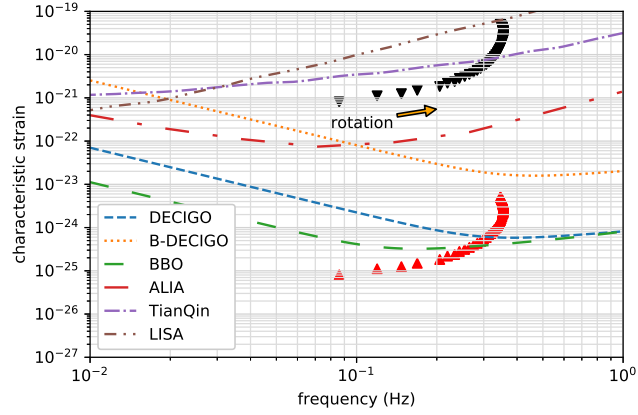}
\caption{Characteristic strain of gravitational wave from the fastest growing eigenmodes for $T=10^8$K, $\rho_c=10^8\mbox{g cm}^{-3}$,
$\alpha=100$, $\beta=10$, and $R_c=0.4$. 
The (black) triangles pointing downward correspond to the case with the saturation amplitude ($\epsilon_I$) being $10^{-1}$,
while the (red) triangles pointing upward correspond to the case with $\epsilon_I=10^{-5}$.  Distance to the source is assumed
to be 20Mpc. 
Along the sequence, the  direction of the faster rotation (measured by the axis ratio) is indicated by an arrrow.
Also plotted are sensitivity curves for DECIGO (dashed), B-DECIGO (dotted), BBO (long-dashed),  ALIA (long-dash-dotted), 
TianQin (dash-dotted), and LISA (dash-double-dotted).
Data of these sensitivity curves are taken from {\tt http://gwplotter.com} except for DECIGO, B-DECIGO, and BBO.
}
\label{fig: hc-T8rho8Rc0.40}
\end{figure}

In Fig.\ref{fig: hc-T8rho8Rc0.40} the sequences of characteristic strain $h_c$ 
for the fastest growing mode in $(T, \rho_c, \alpha, \beta, R_c)=(10^8\mbox{K}, 10^8\mbox{g cm}^{-3}, 100, 10, 0.4)$
stars are plotted. For all models here the fastest growing mode is of inertial nature. Therefore the non-rotating
limit is not in the plots, since the eigenfrequency vanishes. 
The (black) triangles pointing downward are $h_c$ for which the maximal amplitude of $\epsilon_I=0.1$ are
assumed following \cite{Shibata_etal2002}. The (red) triangles pointing upward correspond to the minimal amplitude estimated
in the same way as Rossby-wave instability \citep{2013MNRAS.430.1988M}, $\epsilon_I=10^{-5}$.
They are compared with the noise curves of a few detectors.
We extracted the noise curve data from {\tt GWplotter} ({\tt http://gwplotter.com}) developed by \cite{2015CQGra..32a5014M}
in which ALIA data is taken from \cite{Bender_2013}, 
LISA noise data is from \cite{amaroseoane2012elisa}, and TianQin data is from \cite{Luo_2016}.
For DECIGO and BBO noise data we use an analytic expression from \cite{PhysRevD.83.044011}, while we use an
expression from \cite{10.1093/ptep/pty078} for B-DECIGO.
The remnant mass is in the range of $1.17-1.27M_\odot$ with the core mass fraction $0.65-0.92$. The direction of the sequence
to which the model spin up (as measured by the axis ratio) is indicated by an arrow.
As the remnant rotationally flattens more,
the frequency of the gravitational wave increases and so does the characteristic amplitudes. Near the mass-shedding limit
the frequency decreases (as is seen in Fig.\ref{fig: T7rho8Rc0403} the eigenfrequency has its maximum
before it reaches the mass-shedding limit) though the amplitude increases. 
We see DECIGO and BBO will see the mode for the saturation amplitude of $\epsilon_I={\cal O}[10^{-2} - 10^{-1}]$ 
as is found in \cite{Shibata_etal2002}. For this parameter set, TianQin may see the signal if the remnants is close
to mass-shedding limit and the amplitude of the mode is maximal.
On the other hand the amplitude suggested by the saturation of Rossby-wave instability
may allow the signals to be detected by DECIGO and BBO only for the remnants close to mass-shedding limit.

\begin{figure*}[htpb]
		\gridline{\fig{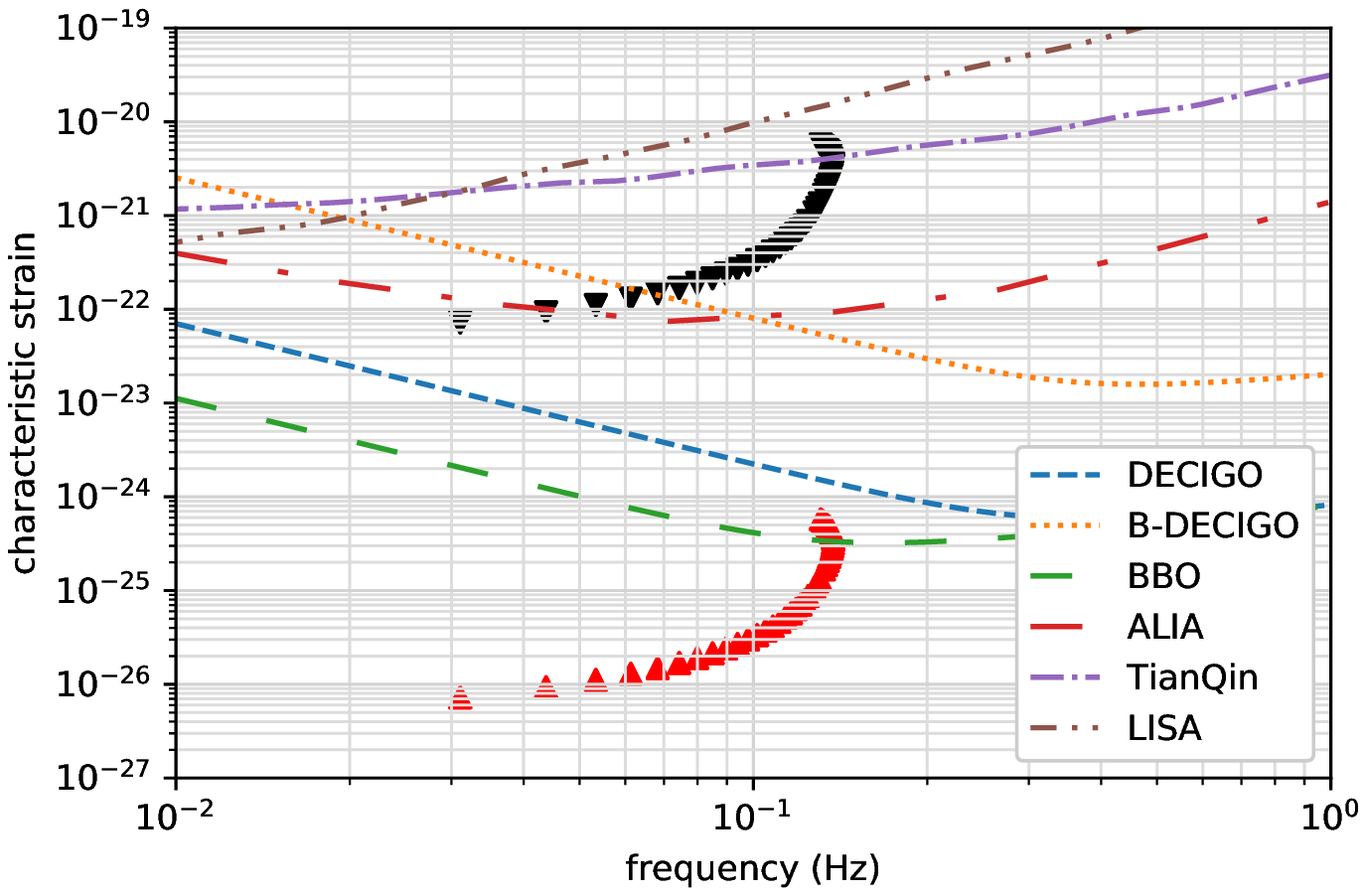}{0.5\textwidth}{(a)}
			\fig{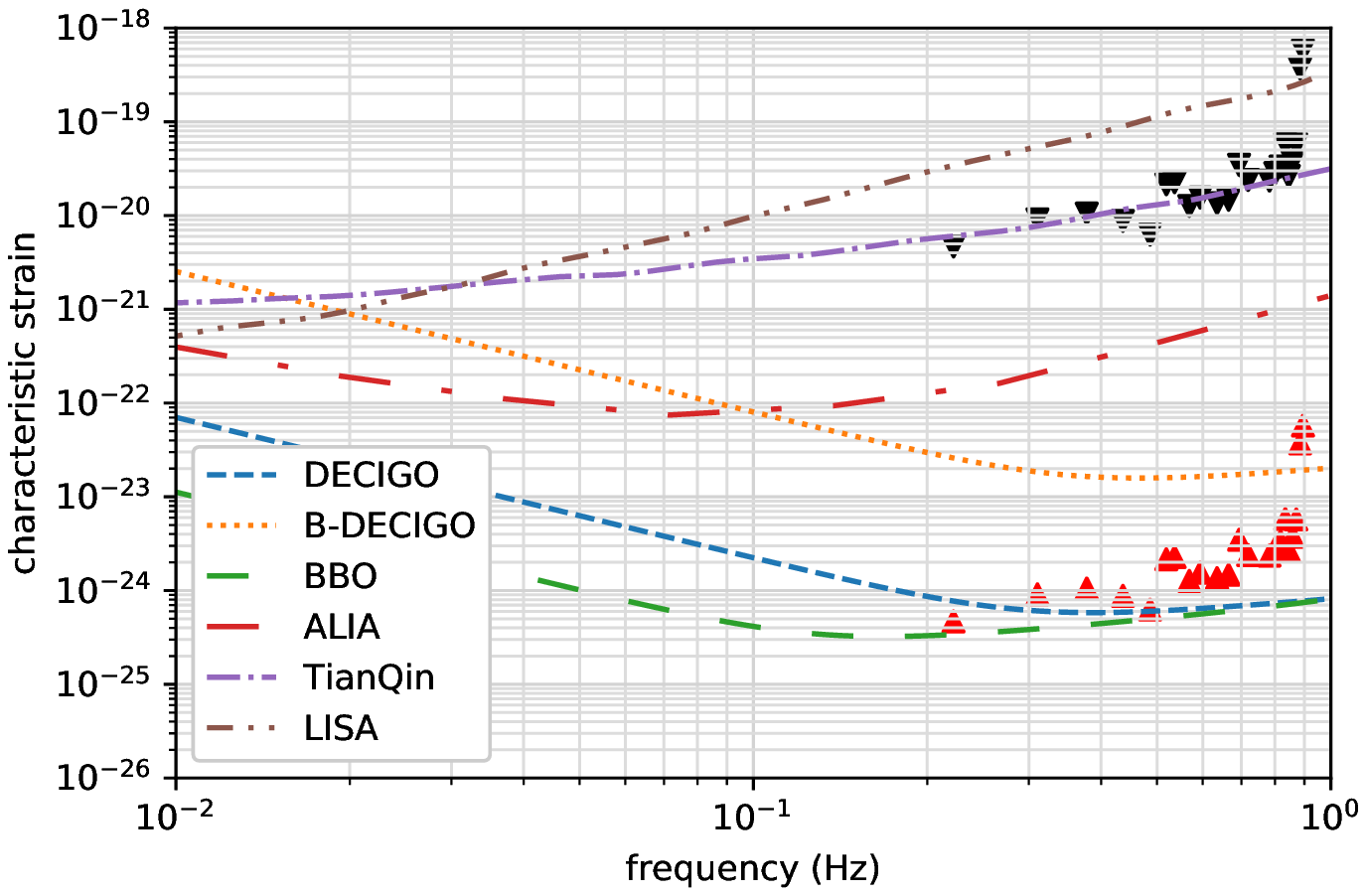}{0.5\textwidth}{(b)}
			}
		\gridline{\fig{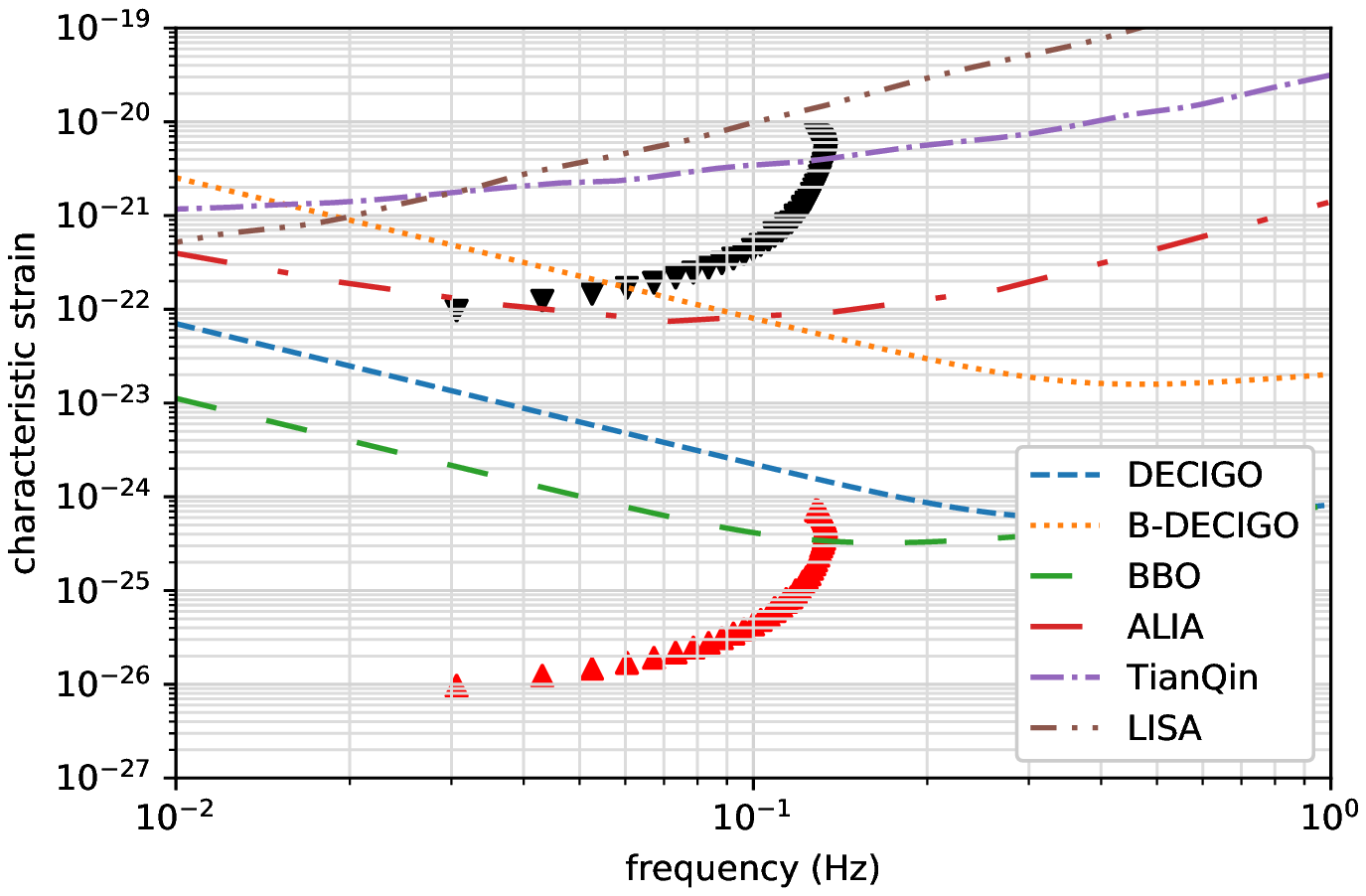}{0.5\textwidth}{(c)}
			\fig{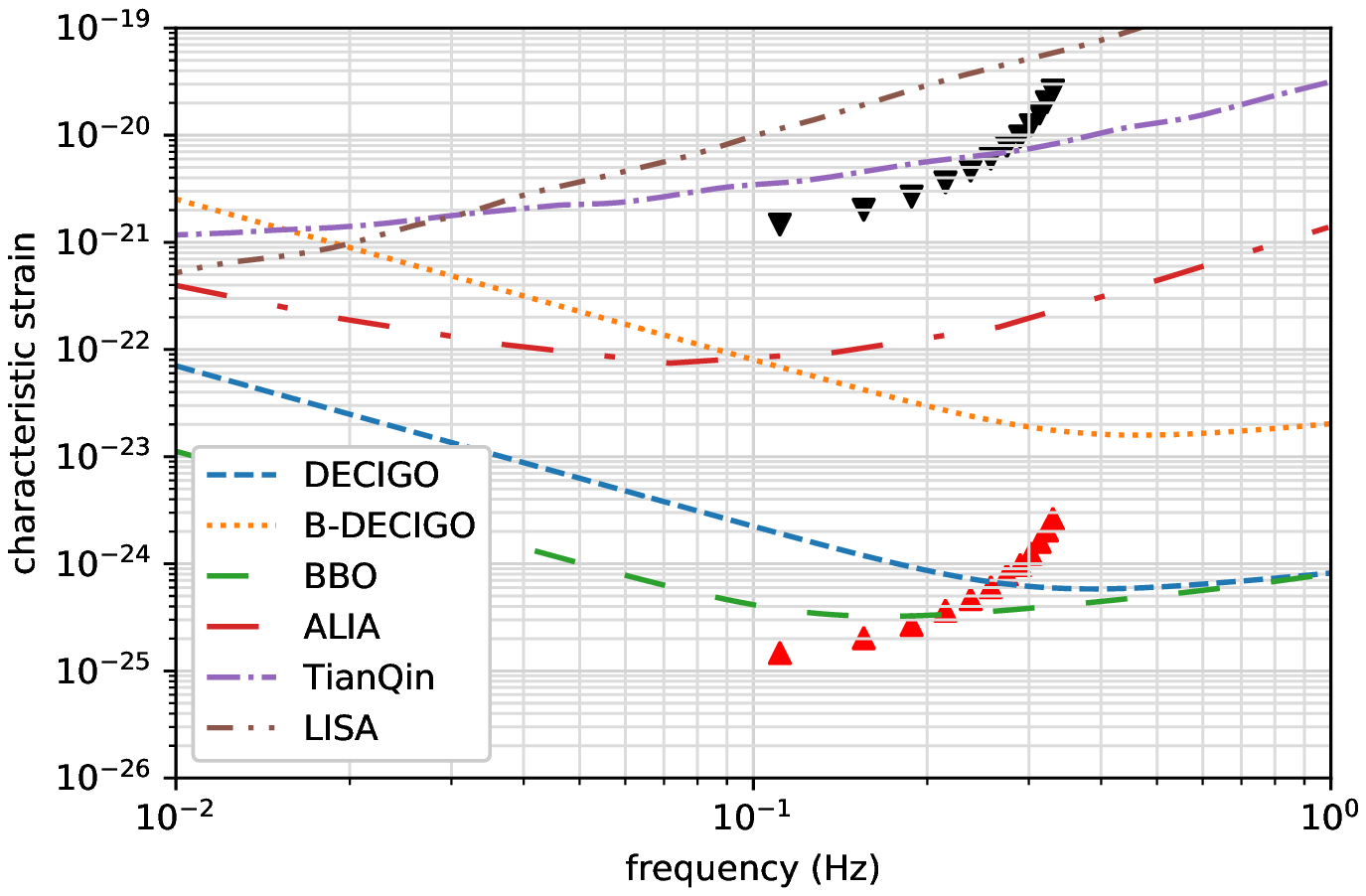}{0.5\textwidth}{(d)}
			}
	\caption{Characteristic strain of GW from unstable oscillations as 
	a function of frequency. 
(a): $(T_c, \rho_c, \alpha, \beta, R_c)=(10^8, 10^7, 100, 10, 0.4)$.
(b): $(T_c, \rho_c, \alpha, \beta, R_c)=(10^8, 10^9, 100, 10, 0.4)$.
(c): $(T_c, \rho_c, \alpha, \beta, R_c)=(10^9, 10^7, 100, 10, 0.4)$.
(d): $(T_c, \rho_c, \alpha, \beta, R_c)=(10^7, 10^8, 20, 10, 0.4)$.
	\label{fig: hc-comparison}
	}
\end{figure*}


In Fig.\ref{fig: hc-comparison} the qualitative behavior of the gravitational wave frequency and its strain is similar
to Fig.\ref{fig: hc-T8rho8Rc0.40}. For less/more massive stars with the lower/higher central density 
the frequency and the strain of the gravitational wave are lower/higher (compare panel (a) and (b) with Fig.\ref{fig: hc-T8rho8Rc0.40}). 
For $\rho=10^9\mbox{g cm}^{-3}$
even the minimal saturation amplitude models may be detectable by BBO or DECIGO, although
the mass range is between $1.35-1.42M_\odot$ and the carbon detonation may occur 
in the early stage of the remnant evolution.
Panel (c) is for a higher temperature than in (b) but with smaller $\rho_c$. 
Therefore the electron degeneracy is weaker. It
increases the strain compared with (a), though the difference is not significant.

For an $\Omega$ profile with less gradient at the core-envelope boundary as in panel (b)
of Fig.\ref{fig: modeT7rho8Rc0.40_bet10set}, the instability becomes weaker in a sense that
the range of axis ratio in which instability occurs is smaller and that the growth time scale
of the fastest growing eigenmode is longer. Still the characteristic strain for this case 
may be large enough to be observed (See panel (d). It corresponds to panel (b) of 
Fig.\ref{fig: modeT7rho8Rc0.40_bet10set}).


\section{Summary and Discussions}
SNeIa have little been considered as sources of GW (see \cite{2011PhRvL.106t1103F} 
for a model where SN Ia explosion of non DWD merger origin as a GW source), though their possible
progenitor binaries may be observed by the space-borne interferometers.
We here consider the possibility of observing merger remnants of DWD
by gravitational wave detectors in decihertz band. A merger
may leave a rapidly and differentially rotating remnant which
have a hot envelope formed through accretion of the secondary
star in the progenitor binary onto the primary. The angular velocity
profile of the remnant in the early phase of its evolution
 is expected to have a steep gradient around
the core-envelope interface. In these system a dynamical instability
is expected which is related to corotation of the oscillation pattern of fluid
to the remnant's rotation. We see the fastest
growing eigenmode has the nature of inertial oscillation, whose
restoring force is the Coriolis force. This is consistent with
the recent result of a study in the context of neutron star
merger \citep{2020arXiv200310198P}, which is reasonable since
the merger process may lead to a similar rotational profile (slowly
and uniformly rotating core surrounded by a rapidly rotating
envelope). The instability mechanism may not be the same as
that in the so-called 'low-T/W' instability found in differentially
rotating stars with a monotonic rotational profile, where it is likely that
overeflection of acoustic wave at the corotation singularity
plays a key role \citep{2017MNRAS.466..600Y}.
It may be possible that overrefrection of the inertial (Rossby) wave 
at the corotation radius plays a similar role in the current case
as in the conventional low-T/W instability.

Saturation of amplitude
of the fastest growing modes may sustain quasi-stable quadrupole deformation
of the remnant. It may emit gravitational wave at nearly constant
frequency, though it only lasts for $10^3-10^4$s in which viscosity
smooth out the differential rotation. Therefore we may see a transient
gravitational wave signal of nearly constant frequency from an unstable
inertial oscillation whose frequency is in decihertz range.
The characteristic strain amplitude at Virgo cluster is well above the noise
curves of such proposed detectors as BBO or DECIGO,
if the density (or mass) and the saturation
amplitude is large enough. It may leads to as much as ten observable
events per year (See Sec.\ref{sec:introduction}). It should be noted
that the present mechanism of gravitational radiation works not only
in a massive merger that may eventually lead to SNIa, but in a lighter
one which survives the thermonuclear runaway. In fact the gravitational
wave from a very massive merger may not be observable if it promptly explodes.

As a final remark, we may speculate a prompt mechanism of SNIa related
to the unstable oscillations. Relatively massive DWD may survive the initial
dynamical accretion phase of its merger and settles down to a quasi-equilibrium
state. If it has a sufficient degree of differential rotation, the instability considered
here grows. The remnant is like a sound box of trapped inertial/sound wave whose 
amplitude grows. The density fluctuation at the core-envelope boundary
may become large enough to trigger deflagration/detonation wave to travel
in the degenerate core, leading to the thermonuclear runaway. If this is the case
the explosion is expected to occur during which the gravitational signal
considered in the current study is observed. This may be an interesting target
of multi-messenger astronomy related to SNIa. It should be noticed that
this mechanism is expected to leave a trace of quadrupolar asymmetry
in the explosion.

\acknowledgments
I thank the anonymous reviewer for her/his useful comments to improve the paper. 
This work is supported by MEXT/JSPS KAKENHI Grant-in-Aid for Scientific Research (C) 18K03641.


\bibliography{bwremgw}{}

\begin{thebibliography}{}
\expandafter\ifx\csname natexlab\endcsname\relax\def\natexlab#1{#1}\fi

\bibitem[{{Abou El-Neaj} {et~al.}(2019){Abou El-Neaj}, {Alpigiani},
  {Amairi-Pyka}, {Araujo}, {Balaz}, {Bassi}, {Bathe-Peters}, {Battelier},
  {Belic}, {Bentine}, {Bernabeu}, {Bertoldi}, {Bingham}, {Blas}, {Bolpasi},
  {Bongs}, {Bose}, {Bouyer}, {Bowcock}, {Bowden}, {Buchmueller}, {Burrage},
  {Calmet}, {Canuel}, {Caramete}, {Carroll}, {Cella}, {Charmandaris},
  {Chattopadhyay}, {Chen}, {Chiofalo}, {Coleman}, {Cotter}, {Cui},
  {Derevianko}, {De Roeck}, {Djordjevic}, {Dornan}, {Doser}, {Drougkakis},
  {Dunningham}, {Dutan}, {Easo}, {Elertas}, {Ellis}, {El Sawy}, {Fassi},
  {Felea}, {Feng}, {Flack}, {Foot}, {Fuentes}, {Gaaloul}, {Gauguet}, {Geiger},
  {Gibson}, {Giudice}, {Goldwin}, {Grachov}, {Graham}, {Grasso}, {van der
  Grinten}, {Gundogan}, {Haehnelt}, {Harte}, {Hees}, {Hobson}, {Holst},
  {Hogan}, {Kasevich}, {Kavanagh}, {von Klitzing}, {Kovachy}, {Krikler},
  {Krutzik}, {Lewicki}, {Lien}, {Liu}, {Gaetano Luciano}, {Magnon}, {Mahmoud},
  {Malik}, {McCabe}, {Mitchell}, {Pahl}, {Pal}, {Pand ey}, {Papazoglou},
  {Paternostro}, {Penning}, {Peters}, {Prevedelli}, {Puthiya-Veettil},
  {Quenby}, {Rasel}, {Ravenhall}, {Rejeb Sfar}, {Ringwood}, {Roura},
  {Sabulsky}, {Sameed}, {Sauer}, {Alaric Schaffer}, {Schiller}, {Schkolnik},
  {Schlippert}, {Schubert}, {Shayeghi}, {Shipsey}, {Signorini},
  {Soares-Santos}, {Sorrentino}, {Singh}, {Sumner}, {Tassis}, {Tentindo},
  {Tino}, {Tinsley}, {Unwin}, {Valenzuela}, {Vasilakis}, {Vaskonen}, {Vogt},
  {Webber-Date}, {Wenzlawski}, {Windpassinger}, {Woltmann}, {Holynski},
  {Yazgan}, {Zhan}, {Zou}, \& {Zupan}}]{2019arXiv190800802A}
{Abou El-Neaj}, Y., {Alpigiani}, C., {Amairi-Pyka}, S., {et~al.} 2019, arXiv
  e-prints, arXiv:1908.00802

\bibitem[{Amaro-Seoane {et~al.}(2012)Amaro-Seoane, Aoudia, Babak, Binetruy,
  Berti, Boh\'{e}, Caprini, Colpi, Cornish, Danzmann, Dufaux, Gair, Jennrich,
  Jetzer, Klein, Lang, Lobo, Littenberg, McWilliams, Nelemans, Petiteau,
  Porter, Schutz, Sesana, Stebbins, Sumner, Vallisneri, Vitale, Volonteri, \&
  Ward}]{amaroseoane2012elisa}
Amaro-Seoane, P., Aoudia, S., Babak, S., {et~al.} 2012, arXiv:1201.3621

\bibitem[{Ando {et~al.}(2010)Ando, Ishidoshiro, Yamamoto, Yagi, Kokuyama,
  Tsubono, \& Takamori}]{PhysRevLett.105.161101}
Ando, M., Ishidoshiro, K., Yamamoto, K., {et~al.} 2010, Phys. Rev. Lett., 105,
  161101

\bibitem[{{Badurina} {et~al.}(2020){Badurina}, {Bentine}, {Blas}, {Bongs},
  {Bortoletto}, {Bowcock}, {Bridges}, {Bowden}, {Buchmueller}, {Burrage},
  {Coleman}, {Elertas}, {Ellis}, {Foot}, {Gibson}, {Haehnelt}, {Harte},
  {Hedges}, {Hobson}, {Holynski}, {Jones}, {Langlois}, {Lellouch}, {Lewicki},
  {Maiolino}, {Majewski}, {Malik}, {March-Russell}, {McCabe}, {Newbold},
  {Sauer}, {Schneider}, {Shipsey}, {Singh}, {Uchida}, {Valenzuela}, {van der
  Grinten}, {Vaskonen}, {Vossebeld}, {Weatherill}, \&
  {Wilmut}}]{2020JCAP...05..011B}
{Badurina}, L., {Bentine}, E., {Blas}, D., {et~al.} 2020, \jcap, 2020, 011

\bibitem[{{Balbus} \& {Hawley}(1991)}]{Balbus-Hawley1991a}
{Balbus}, S.~A., \& {Hawley}, J.~F. 1991, \apj, 376, 214

\bibitem[{Bender {et~al.}(2013)Bender, Begelman, \& Gair}]{Bender_2013}
Bender, P.~L., Begelman, M.~C., \& Gair, J.~R. 2013, Classical and Quantum
  Gravity, 30, 165017

\bibitem[{{Benz} {et~al.}(1990){Benz}, {Bowers}, {Cameron}, \&
  {Press}}]{Benz1990}
{Benz}, W., {Bowers}, R.~L., {Cameron}, A.~G.~W., \& {Press}, W.~H.~. 1990,
  \apj, 348, 647

\bibitem[{{Blanchet} {et~al.}(1990){Blanchet}, {Damour}, \&
  {Schaefer}}]{1990MNRAS.242..289B}
{Blanchet}, L., {Damour}, T., \& {Schaefer}, G. 1990, \mnras, 242, 289

\bibitem[{{Centrella} {et~al.}(2001){Centrella}, {New}, {Lowe}, \&
  {Brown}}]{Centrella_etal2001}
{Centrella}, J.~M., {New}, K.~C.~B., {Lowe}, L.~L., \& {Brown}, J.~D. 2001,
  \apjl, 550, L193

\bibitem[{{Chandrasekhar}(1960)}]{1960PNAS...46..253C}
{Chandrasekhar}, S. 1960, Proceedings of the National Academy of Science, 46,
  253

\bibitem[{{Chandrasekhar}(1987)}]{ChandraEFE}
---. 1987, {Ellipsoidal figures of equilibrium} (Dover, New York)

\bibitem[{{Corvino} {et~al.}(2010){Corvino}, {Rezzolla}, {Bernuzzi}, {De
  Pietri}, \& {Giacomazzo}}]{Corvino_etal2010}
{Corvino}, G., {Rezzolla}, L., {Bernuzzi}, S., {De Pietri}, R., \&
  {Giacomazzo}, B. 2010, Classical and Quantum Gravity, 27, 114104

\bibitem[{{Cox} \& {Giuli}(1968)}]{1968pss..book.....C}
{Cox}, J.~P., \& {Giuli}, R.~T. 1968, {Principles of stellar structure}

\bibitem[{Crowder \& Cornish(2005)}]{PhysRevD.72.083005}
Crowder, J., \& Cornish, N.~J. 2005, Phys. Rev. D, 72, 083005

\bibitem[{{Dan} {et~al.}(2011){Dan}, {Rosswog}, {Guillochon}, \&
  {Ramirez-Ruiz}}]{Dan2011}
{Dan}, M., {Rosswog}, S., {Guillochon}, J., \& {Ramirez-Ruiz}, E. 2011, \apj,
  737, 89

\bibitem[{{Dan} {et~al.}(2012){Dan}, {Rosswog}, {Guillochon}, \&
  {Ramirez-Ruiz}}]{Dan2012}
---. 2012, \mnras, 422, 2417

\bibitem[{{Falta} {et~al.}(2011){Falta}, {Fisher}, \&
  {Khanna}}]{2011PhRvL.106t1103F}
{Falta}, D., {Fisher}, R., \& {Khanna}, G. 2011, \prl, 106, 201103

\bibitem[{{Ferrario} {et~al.}(2015){Ferrario}, {de Martino}, \&
  {G{\"a}nsicke}}]{2015SSRv..191..111F}
{Ferrario}, L., {de Martino}, D., \& {G{\"a}nsicke}, B.~T. 2015, \ssr, 191, 111

\bibitem[{{Graham} {et~al.}(2017){Graham}, {Hogan}, {Kasevich}, {Rajendran}, \&
  {Romani}}]{2017arXiv171102225G}
{Graham}, P.~W., {Hogan}, J.~M., {Kasevich}, M.~A., {Rajendran}, S., \&
  {Romani}, R.~W. 2017, arXiv e-prints, arXiv:1711.02225

\bibitem[{{Hachisu}(1986)}]{1986ApJS...61..479H}
{Hachisu}, I. 1986, \apjs, 61, 479

\bibitem[{Huang {et~al.}(2020)Huang, Hu, Korol, Li, Liang, Lu, Wang, Yu, \&
  Mei}]{PhysRevD.102.063021}
Huang, S.-J., Hu, Y.-M., Korol, V., {et~al.} 2020, Phys. Rev. D, 102, 063021

\bibitem[{{Iben} \& {Tutukov}(1984)}]{1984ApJS...54..335I}
{Iben}, I., J., \& {Tutukov}, A.~V. 1984, \apjs, 54, 335

\bibitem[{{Ipser} \& {Lindblom}(1991)}]{1991ApJ...379..285I}
{Ipser}, J.~R., \& {Lindblom}, L. 1991, \apj, 379, 285

\bibitem[{Isoyama {et~al.}(2018)Isoyama, Nakano, \&
  Nakamura}]{10.1093/ptep/pty078}
Isoyama, S., Nakano, H., \& Nakamura, T. 2018, Progress of Theoretical and
  Experimental Physics, 2018,
  https://academic.oup.com/ptep/article-pdf/2018/7/073E01/25332865/pty078.pdf,
  073E01

\bibitem[{{Kalogera} {et~al.}(2001){Kalogera}, {Narayan}, {Spergel}, \&
  {Taylor}}]{Kalogera2001}
{Kalogera}, V., {Narayan}, R., {Spergel}, D.~N., \& {Taylor}, J.~H. 2001, \apj,
  556, 340

\bibitem[{{Kashyap} {et~al.}(2017){Kashyap}, {Fisher}, {Garc{\'{\i}}a-Berro},
  {Aznar-Sigu{\'a}n}, {Ji}, \& {Lor{\'e}n-Aguilar}}]{Kashyap2017}
{Kashyap}, R., {Fisher}, R., {Garc{\'{\i}}a-Berro}, E., {et~al.} 2017, \apj,
  840, 16

\bibitem[{Kawamura {et~al.}(2006)Kawamura, Nakamura, Ando, Seto, Tsubono,
  Numata, Takahashi, Nagano, Ishikawa, Musha, ichi Ueda, Sato, Hosokawa,
  Agatsuma, Akutsu, suke Aoyanagi, Arai, Araya, Asada, Aso, Chiba, Ebisuzaki,
  Eriguchi, Fujimoto, Fukushima, Futamase, Ganzu, Harada, Hashimoto, Hayama,
  Hikida, Himemoto, Hirabayashi, Hiramatsu, Ichiki, Ikegami, Inoue, Ioka,
  Ishidoshiro, Itoh, Kamagasako, Kanda, Kawashima, Kirihara, Kiuchi, Kobayashi,
  Kohri, Kojima, Kokeyama, Kozai, Kudoh, Kunimori, Kuroda, ichi Maeda,
  Matsuhara, Mino, Miyakawa, Miyoki, Mizusawa, Morisawa, Mukohyama, Naito,
  Nakagawa, Nakamura, Nakano, Nakao, Nishizawa, Niwa, Nozawa, Ohashi, Ohishi,
  Ohkawa, Okutomi, Oohara, Sago, Saijo, Sakagami, Sakata, Sasaki, Sato,
  Shibata, Shinkai, Somiya, Sotani, Sugiyama, Tagoshi, Takahashi, Takahashi,
  Takahashi, Takano, Tanaka, Taniguchi, Taruya, Tashiro, Tokunari, Tsujikawa,
  Tsunesada, Yamamoto, Yamazaki, Yokoyama, Yoo, Yoshida, \&
  Yoshino}]{Kawamura_2006}
Kawamura, S., Nakamura, T., Ando, M., {et~al.} 2006, Classical and Quantum
  Gravity, 23, S125

\bibitem[{{Licquia} \& {Newman}(2015)}]{2015ApJ...806...96L}
{Licquia}, T.~C., \& {Newman}, J.~A. 2015, \apj, 806, 96

\bibitem[{{Lor{\'e}n-Aguilar} {et~al.}(2005){Lor{\'e}n-Aguilar}, {Guerrero},
  {Isern}, {Lobo}, \& {Garc{\'{\i}}a-Berro}}]{Loren-Aguilar2005}
{Lor{\'e}n-Aguilar}, P., {Guerrero}, J., {Isern}, J., {Lobo}, J.~A., \&
  {Garc{\'{\i}}a-Berro}, E. 2005, \mnras, 356, 627

\bibitem[{{Lor{\'e}n-Aguilar} {et~al.}(2009){Lor{\'e}n-Aguilar}, {Isern}, \&
  {Garc{\'{\i}}a-Berro}}]{Loren-Aguilar2009}
{Lor{\'e}n-Aguilar}, P., {Isern}, J., \& {Garc{\'{\i}}a-Berro}, E. 2009, \aap,
  500, 1193

\bibitem[{{Lovelace} {et~al.}(2009){Lovelace}, {Turner}, \&
  {Romanova}}]{2009ApJ...701..225L}
{Lovelace}, R.~V.~E., {Turner}, L., \& {Romanova}, M.~M. 2009, \apj, 701, 225

\bibitem[{Luo {et~al.}(2016)Luo, Chen, Duan, Gong, Hu, Ji, Liu, Mei, Milyukov,
  Sazhin, Shao, Toth, Tu, Wang, Wang, Yeh, Zhan, Zhang, Zharov, \&
  Zhou}]{Luo_2016}
Luo, J., Chen, L.-S., Duan, H.-Z., {et~al.} 2016, Classical and Quantum
  Gravity, 33, 035010

\bibitem[{{Maoz} {et~al.}(2018){Maoz}, {Hallakoun}, \& {Badenes}}]{Maoz2018}
{Maoz}, D., {Hallakoun}, N., \& {Badenes}, C. 2018, \mnras, 476, 2584

\bibitem[{{Meheut} {et~al.}(2013){Meheut}, {Lovelace}, \&
  {Lai}}]{2013MNRAS.430.1988M}
{Meheut}, H., {Lovelace}, R.~V.~E., \& {Lai}, D. 2013, \mnras, 430, 1988

\bibitem[{{Moore} {et~al.}(2015){Moore}, {Cole}, \&
  {Berry}}]{2015CQGra..32a5014M}
{Moore}, C.~J., {Cole}, R.~H., \& {Berry}, C.~P.~L. 2015, Classical and Quantum
  Gravity, 32, 015014

\bibitem[{{Oohara} \& {Nakamura}(1989)}]{1989PThPh..82..535O}
{Oohara}, K., \& {Nakamura}, T. 1989, Progress of Theoretical Physics, 82, 535

\bibitem[{{Ostriker} \& {Bodenheimer}(1973)}]{Ostriker_Bodenheimer1973}
{Ostriker}, J.~P., \& {Bodenheimer}, P. 1973, \apj, 180, 171

\bibitem[{{Ostriker} \& {Tassoul}(1969)}]{Ostriker_Tassoul1969}
{Ostriker}, J.~P., \& {Tassoul}, J.~L. 1969, \apj, 155, 987

\bibitem[{{Ou} \& {Tohline}(2006)}]{Ou_Tohline2006}
{Ou}, S., \& {Tohline}, J.~E. 2006, \apj, 651, 1068

\bibitem[{{Pakmor} {et~al.}(2012){Pakmor}, {Kromer}, {Taubenberger}, {Sim},
  {R{\"o}pke}, \& {Hillebrandt}}]{Pakmor2012}
{Pakmor}, R., {Kromer}, M., {Taubenberger}, S., {et~al.} 2012, \apjl, 747, L10

\bibitem[{{Passamonti} \& {Andersson}(2015)}]{Passamonti_Andersson2015}
{Passamonti}, A., \& {Andersson}, N. 2015, \mnras, 446, 555

\bibitem[{{Passamonti} \& {Andersson}(2020)}]{2020arXiv200310198P}
---. 2020, arXiv e-prints, arXiv:2003.10198

\bibitem[{{Pickett} {et~al.}(1996){Pickett}, {Durisen}, \&
  {Davis}}]{Picket_etal1996}
{Pickett}, B.~K., {Durisen}, R.~H., \& {Davis}, G.~A. 1996, \apj, 458, 714

\bibitem[{{Piro}(2008)}]{2008ApJ...679..616P}
{Piro}, A.~L. 2008, \apj, 679, 616

\bibitem[{{Press} {et~al.}(1992){Press}, {Teukolsky}, {Vetterling}, \&
  {Flannery}}]{1992nrfa.book.....P}
{Press}, W.~H., {Teukolsky}, S.~A., {Vetterling}, W.~T., \& {Flannery}, B.~P.
  1992, {Numerical recipes in FORTRAN. The art of scientific computing}

\bibitem[{{Raskin} {et~al.}(2012){Raskin}, {Scannapieco}, {Fryer},
  {Rockefeller}, \& {Timmes}}]{Raskin2012}
{Raskin}, C., {Scannapieco}, E., {Fryer}, C., {Rockefeller}, G., \& {Timmes},
  F.~X. 2012, \apj, 746, 62

\bibitem[{{Saijo} {et~al.}(2003){Saijo}, {Baumgarte}, \&
  {Shapiro}}]{Saijo_etal2003}
{Saijo}, M., {Baumgarte}, T.~W., \& {Shapiro}, S.~L. 2003, \apj, 595, 352

\bibitem[{{Saijo} \& {Yoshida}(2006)}]{Saijo_Yoshida2006}
{Saijo}, M., \& {Yoshida}, S. 2006, \mnras, 368, 1429

\bibitem[{{Saijo} \& {Yoshida}(2016)}]{2016PhRvD..94h4032S}
---. 2016, \prd, 94, 084032

\bibitem[{{Sato} {et~al.}(2015){Sato}, {Nakasato}, {Tanikawa}, {Nomoto},
  {Maeda}, \& {Hachisu}}]{Sato2015}
{Sato}, Y., {Nakasato}, N., {Tanikawa}, A., {et~al.} 2015, \apj, 807, 105

\bibitem[{{Sato} {et~al.}(2016){Sato}, {Nakasato}, {Tanikawa}, {Nomoto},
  {Maeda}, \& {Hachisu}}]{Sato2016}
---. 2016, \apj, 821, 67

\bibitem[{{Schutz}(1980)}]{Schutz1980_pap3}
{Schutz}, B.~F. 1980, \mnras, 190, 21

\bibitem[{Sedda {et~al.}(2019)Sedda, Berry, Jani, Amaro-Seoane, Auclair, Baird,
  Baker, Berti, Breivik, Burrows, Caprini, Chen, Doneva, Ezquiaga, Ford, Katz,
  Kolkowitz, McKernan, Mueller, Nardini, Pikovski, Rajendran, Sesana, Shao,
  Tamanini, Vartanyan, Warburton, Witek, Wong, \& Zevin}]{sedda2019missing}
Sedda, M.~A., Berry, C. P.~L., Jani, K., {et~al.} 2019, The Missing Link in
  Gravitational-Wave Astronomy: Discoveries waiting in the decihertz range, , ,
  arXiv:1908.11375

\bibitem[{{Segretain} {et~al.}(1997){Segretain}, {Chabrier}, \&
  {Mochkovitch}}]{Segretain1997}
{Segretain}, L., {Chabrier}, G., \& {Mochkovitch}, R. 1997, \apj, 481, 355

\bibitem[{{Shakura} \& {Sunyaev}(1973)}]{1973A&A....24..337S}
{Shakura}, N.~I., \& {Sunyaev}, R.~A. 1973, \aap, 500, 33

\bibitem[{{Shen} {et~al.}(2012){Shen}, {Bildsten}, {Kasen}, \&
  {Quataert}}]{Shen2012}
{Shen}, K.~J., {Bildsten}, L., {Kasen}, D., \& {Quataert}, E. 2012, \apj, 748,
  35

\bibitem[{{Shibata} {et~al.}(2002){Shibata}, {Karino}, \&
  {Eriguchi}}]{Shibata_etal2002}
{Shibata}, M., {Karino}, S., \& {Eriguchi}, Y. 2002, \mnras, 334, L27

\bibitem[{{Shibata} {et~al.}(2003){Shibata}, {Karino}, \&
  {Eriguchi}}]{Shibata_etal2003}
---. 2003, \mnras, 343, 619

\bibitem[{{Spruit}(1999)}]{1999A&A...349..189S}
{Spruit}, H.~C. 1999, \aap, 349, 189

\bibitem[{{Spruit}(2002)}]{2002A&A...381..923S}
---. 2002, \aap, 381, 923

\bibitem[{{Tanikawa} {et~al.}(2015){Tanikawa}, {Nakasato}, {Sato}, {Nomoto},
  {Maeda}, \& {Hachisu}}]{Tanikawa2015}
{Tanikawa}, A., {Nakasato}, N., {Sato}, Y., {et~al.} 2015, \apj, 807, 40

\bibitem[{{Tassoul}(1978{\natexlab{a}})}]{1978trs..book.....T}
{Tassoul}, J.-L. 1978{\natexlab{a}}, {Theory of rotating stars}

\bibitem[{{Tassoul}(1978{\natexlab{b}})}]{Tassoul1978book}
---. 1978{\natexlab{b}}, {Theory of rotating stars, Chap.10} (Princeton
  University Press, New Jersey)

\bibitem[{{Tassoul} \& {Ostriker}(1968)}]{Tassoul_Ostriker1968}
{Tassoul}, J.~L., \& {Ostriker}, J.~P. 1968, \apj, 154, 613

\bibitem[{{Tayler}(1973)}]{Tayler1973}
{Tayler}, R.~J. 1973, \mnras, 161, 365

\bibitem[{{Velikhov}(1959)}]{Velikhov1959}
{Velikhov}, E.~P. 1959, Zhur. Eksptl'. i Teoret. Fiz., 36

\bibitem[{{Watts} {et~al.}(2005){Watts}, {Andersson}, \&
  {Jones}}]{Watts_etal2005}
{Watts}, A.~L., {Andersson}, N., \& {Jones}, D.~I. 2005, \apjl, 618, L37

\bibitem[{{Webbink}(1984)}]{1984ApJ...277..355W}
{Webbink}, R.~F. 1984, \apj, 277, 355

\bibitem[{{Xie} {et~al.}(2020){Xie}, {Hawke}, {Passamonti}, \&
  {Andersson}}]{2020PhRvD.102d4040X}
{Xie}, X., {Hawke}, I., {Passamonti}, A., \& {Andersson}, N. 2020, \prd, 102,
  044040

\bibitem[{Yagi \& Seto(2011)}]{PhysRevD.83.044011}
Yagi, K., \& Seto, N. 2011, Phys. Rev. D, 83, 044011

\bibitem[{{Yoon} {et~al.}(2007{\natexlab{a}}){Yoon}, {Podsiadlowski}, \&
  {Rosswog}}]{Yoon2007}
{Yoon}, S.-C., {Podsiadlowski}, P., \& {Rosswog}, S. 2007{\natexlab{a}},
  \mnras, 380, 933

\bibitem[{{Yoon} {et~al.}(2007{\natexlab{b}}){Yoon}, {Podsiadlowski}, \&
  {Rosswog}}]{2007MNRAS.380..933Y}
{Yoon}, S.~C., {Podsiadlowski}, P., \& {Rosswog}, S. 2007{\natexlab{b}},
  \mnras, 380, 933

\bibitem[{{Yoshida}(2019)}]{2019MNRAS.486.2982Y}
{Yoshida}, S. 2019, \mnras, 486, 2982

\bibitem[{{Yoshida} \& {Saijo}(2017)}]{2017MNRAS.466..600Y}
{Yoshida}, S., \& {Saijo}, M. 2017, \mnras, 466, 600

\bibitem[{{Zwerger} \& {M\"uller}(1997)}]{1997A&A...320..209Z}
{Zwerger}, T., \& {M\"uller}, E. 1997, \aap, 320, 209

\end{thebibliography}
\bibliographystyle{aasjournal}



\end{document}